\documentclass[a4paper]{article}
\usepackage{authblk}
\usepackage{amsmath}
\usepackage{amssymb}
\usepackage{graphicx}

\def\ket#1{|{#1}\rangle}
\def\bra#1{\langle {#1} |}
\def\scalar#1#2{\langle{#1}|{#2}\rangle}
\def\be{\begin{equation}}
\def\ee{\end{equation}}

\def\XXint#1#2#3{{\setbox0=\hbox{$#1{#2#3}{\int}$}
     \vcenter{\hbox{$#2#3$}}\kern-.5\wd0}}

\DeclareMathOperator{\arctanh}{arctanh}

\title{Numerical and analytical results for the two disks capacitor problem}

\author[1,2]{Giampiero Paffuti}
\affil[1]{Dipartimento di Fisica - Universit\`a di Pisa,  Largo Pontecorvo 3, Pisa, Italy}
\affil[2]{INFN sezione di Pisa, L.go Pontecorvo 3 Ed.  C, I-56127 Pisa, Italy}

\begin{document}
\maketitle

\begin{abstract}
In this paper we study the two disks capacitor, for equal and different radii.
The new results obtained allow a complete characterization of capacity coefficients
and forces at short distances. An extensive numerical calculation confirms the theoretical results.
The study shows the existence of a hierarchy in the divergent behavior of the capacitance coefficients and this implies some unusual 
behavior of the forces, strictly related to the dimensionality of the near-contact zone between electrodes.

\end{abstract}

\section{Introduction}
In this work analytical and numerical tools are integrated to give a complete
characterization of short distance behavior of a two disks capacitor, with arbitrary radii.
On the numerical side it is found that a simple quadrature procedure, properly regularized, is a rather efficient 
 method for the calculation of the capacity. We apply this method improving  the existing results in the literature for the case of equal disks, and producing the first new results in the case of discs of different radii.
The analytical counterpart includes the elaboration of some recent results on different disks and the first calculation of the  
sub-leading terms in short-distance expansion. The interest in this calculation is  not dictated by merely  formal reasons as these terms play an essential role in determining the forces between conductors at small separations. A second reason of interest lies in the search for a classification of divergent terms for capacity coefficients. It has been argued in previous works that at small distances there is a single dominant quantity, while it is possible to define two independent combinations of capacitance coefficients showing a regular behavior in this regime. The case of different radii turns out to be 
the more interesting: the capacitance coefficients can be 
organized in linear combinations which show a hierarchy 
in the regular behavior at short distances and this
 structure is preserved at the sub-leading order. Even more interesting is that these combinations are the same appearing in the study of the electrostatic forces between two conductors. The regularity properties at leading order are common to every couple of conductors, it would be very interesting if the classification of sub-leading terms survives for all system of conductors, as these determine the short distance behavior of forces. 
The paper is organized as follows. In section~\ref{sezdischiuguali} we recall some general theoretical aspects for the problem of two equal disks. 
In section~\ref{eqdisksnum} we present our numerical results and the general guidelines followed for the computations in this paper.
In section~\ref{diffdisksth} we approach the problem of disks with different radii and give a new result for the sub-leading corrections on the behavior at short distances. The results describe completely the capacitance matrix at the level $o(\ell)$, where $\ell$ is the separation between the disks, i.e. we compute all corrections of the form $\ell\log^2\ell, \ell\log\ell$ and the finite terms of order $\ell$.
In section~\ref{diffdisksnum} we present our numerical results on the system. The computation is the first on this problem and is in agreement with the theoretical calculations. The implications of these results for the computation of forces between electrodes are briefly sketched. 

\section{Equal disks: a short review of the problem\label{sezdischiuguali}}

The main scope of this paper is to present novel numeric and analytical results on the coefficients of capacitance 
for a system made of different disks. Since the equations to be analyzed are somewhat a generalization of the well studied case of equal disks it is useful to review the mathematical context and present an efficient numeric approach for this particular configuration.
Let us consider a couple of conducting parallel disks of radius $a$, coaxial and at distance $\ell$. The basic equation for the study of this system is the Love's equation\cite{Love,nick,sne}
\be
1 =  f_L(t) - \int_0^1 K(t,z;\kappa)  f_L(z)\,dz\,.\label{eqaccoppiateug1.3love}\ee
where
\be K(t,z;\kappa) = \frac{\kappa}{\pi} \left(\dfrac1{(z-t)^2 + \kappa^2}
+ \dfrac1{(z + t)^2 + \kappa^2}\right)\,;\qquad \text{with:}\;\kappa = \ell/a\,.
\label{kernelsing}\ee
In the following it will be important the asymptotic solution $f_L$ for $\kappa\to 0$. In the ``bulk region'', i.e. except a small interval of order $\kappa$ near $t=1$, where $f_L$ is finite, we have\cite{Hutson}
\be f_L^{(\kappa)}(t) = 
\left\{ \frac1\kappa \sqrt{1-t^2} + (1-t^2)^{-1/2}\frac1{2\pi}\left(
1 + \log\frac{16\pi}{\kappa} - t \log\frac{1+t}{1-t}\right)\right\} + {o}(1)\,.\label{hutson00}
\ee
We will indicate explicitly, when necessary, the length scale in the function, here $\kappa$.

A simple generalization\cite{sne,Carlson,paf} of \eqref{eqaccoppiateug1.3love}:
\be
V_1 = f_1(t) +\int_0^1 K(t,z;\kappa) f_2(z)\,dz\,,\quad 
V_2 = f_2(t) +  \int_0^1 K(t,z;\kappa) f_1(z)\,dz\,,
\label{eqaccoppiateug1}
\ee
allows the computation of capacitance coefficients $C_{11}, C_{12}$,  defined by the linear system which relates the charges on the conductors to their potentials
\be Q_i= \sum_j C_{ij} V_j\,.\label{capcoeff}\ee 
The charges on the disks are shown to be
\be Q_1 = a\,\frac2\pi \int_0^1 f_1(t)\,dt \,, \qquad Q_2 = a\,\frac2\pi \int_0^1 f_2(t)\,dt\,.\label{cariche1}\ee
For $V_1 = 1, V_2 = 0$ in \eqref{eqaccoppiateug1} we have to solve the system 
\be 1 = f_1(t) +\int_0^1 K(t,z;\kappa) f_2(z)\,dz\,,\quad 
0 = f_2(t) +  \int_0^1 K(t,z;\kappa) f_1(z)\,dz\,,\label{eqaccoppiateug1.2}\ee and expressing the charges through the solutions $f_1, f_2$, we have:
\be C_{11} = C_{22} = a\,\frac2\pi \int_0^1 f_1(t)\,dt \,, \qquad C_{21} = C_{12} = a\,\frac2\pi \int_0^1 f_2(t)\,dt\,.\label{cariche1.3}\ee
It is convenient for theoretical and numerical purposes to use the combinations
\be C = \dfrac{C_{11} - C_{12}}{2}\,,\quad C_{g} = C_{11}+C_{12}\,.\label{defc}\ee
The parameter $C$ is the usual relative capacitance, i.e. the absolute value of the charge on each disk at opposite unity potentials; $C_g$ is the ratio of the charge on either disk with respect to a common potential.  
The corresponding decomposition for the system \eqref{eqaccoppiateug1.2}
\be f_1(t) = \frac12 f(t) + \frac12 g(t)\,;\qquad f_2(t) = -\frac12 f(t) + \frac12 g(t)\,.\label{cambiovariabili}\ee
gives, adding and subtracting the two equations \eqref{eqaccoppiateug1.2}, a decoupled system:
\be
1 = g(t) + \int_0^1 K(t,z;\kappa) g(z)\,dz\,;\quad
1 = f(t) - \int_0^1 K(t,z;\kappa) f(z)\, dz \,.\label{dec1}
\ee
The second equation is the Love equation, i.e. $f= f_L$.
From \eqref{cariche1},\eqref{cariche1.3} we have:
\be C = \frac{a}{\pi}\int_0^1 f_L(t) dt \,; \qquad
C_g = \frac{2a}{\pi}\int_0^1 g(t) dt \,.
\label{eqaccoppiateug1.4lovebis}\ee
The short distance expansions of coefficients $C, C_g$ for equal disks are:
\be C\to C_K + C_K^{(1)} =
a\left\{ \frac{1}{4\kappa} + \frac{1}{4\pi}\left[ \log\left(16\pi \frac 1 \kappa\right) - 1\right]\right\}
+ a\left\{\frac1{16\pi^2} \kappa\left[ \left(\log\frac{\kappa}{16\pi}\right)^2-2\right] \right\}\,,
\label{corrShaw}
\ee
and
\be C_g = a\left[ \frac1\pi + \frac{\kappa}{2\pi^2}\left(\gamma - \log(\kappa)\right)\right]\,.\label{valorecgasint}\ee
The first term in \eqref{corrShaw} is due to the pioneering work of Kirchhoff \cite{Kirchh}. The second term,
a sub-leading correction, has been computed by S.Shaw \cite{Shaw} and improved and corrected in \cite{Wigg,chew}. 
It is worth mentioning 
a different approximation for $\kappa\to 0$, proposed by Ignatowsky\cite{igna}, this approximation differs from $C_K$ in
\eqref{corrShaw} by the substitution $\log(16\pi)-1 \to \log(8)- 1/2$.
The interesting point is that P\'olya and Szeg\"o\cite{polia} showed that Ignatowsky result is a lower bound for the capacitance.

The expansion \eqref{valorecgasint} has been obtained in \cite{paf,mac2}, where the constant $\gamma$ has been estimated using the
preliminary results of the present work as $\gamma\simeq 2.1450(2)\simeq 1 + \log\pi$.

From \eqref{corrShaw} and \eqref{valorecgasint} it follows that $C$ and $C_g$ satisfy the request to classify capacitance coefficients according their different behavior for small distances, as outlined in the introduction.
The first term in \eqref{valorecgasint} reflects the property\cite{paf}
\be \lim_{\kappa\to 0} C_g(\kappa) = \frac{C_T}{2} \label{vallimcg}\ee
where $C_T = 2a/\pi$ is the self-capacity of the system obtained when the two disks collapse, and is finite while $C$ diverges.  
 It is remarkable that also the next order of $C_g$ is less singular than the corresponding term in $C$, i.e. the $\kappa\log^2\kappa$ term is absent.
The same combinations 
\eqref{defc} enter directly in the force between two electrodes. In general the force between two electrodes at distance $\ell$ is given by
\be F = - \frac{\partial}{\partial\ell} \frac12 M_{ij} Q_i Q_j \label{foza1}\ee
where the potential matrix $M_{ij}$ is the inverse of the capacitance matrix $C_{ij}$. 
For two equal conductors with charges $Q_1, Q_2$ it is easy to show\cite{mac2} that the force can be written in the form
\be F(Q_1,Q_2,\kappa) = - \dfrac{(Q_1+Q_2)^2}{4}\frac{\partial}{\partial\ell}
\frac1{C_g} \\
- \dfrac{(Q_1-Q_2)^2}{8}\frac{\partial }{\partial\ell}
\frac1{C}\,.
\label{forza2}\ee
This decomposition is completely general for equal conductors and can be generalized for different conductors.
To avoid misunderstanding it must be stressed that \eqref{forza2} is always valid, but the dependence on $\ell$ is hidden also in the charges $Q_i$ if these charges are not fixed, as in the case of fixed potentials. The last case can be handled using 
\eqref{capcoeff} so in the following we consider the case of fixed charges, unless stated otherwise.

The contact between the electrodes in the limit $\ell\to 0$ depends on the system. Two spheres have a point-like contact, 
two parallel cylinders generally touch each other along a line and two planar electrodes 
 touch each other through a surface. The behavior of the relative capacitance $C$ is completely different in the three cases, being respectively $C_{sphere}\sim 1/\log(\ell)$, $C_{cyl}\sim 1/\ell^{1/2}$, $C_{planar}\sim 1/\ell$.
 This imply that the relative importance of the two terms in \eqref{forza2} depends on the dimensionality of the contact.
 In particular for planar electrodes, the case we are interested in, the second term in \eqref{forza2} gives rise to a constant force.
 The coefficient $C_g$ in the first term has a constant limit for $\ell\to 0$ then its contribution to the force depends on the next order in the short distance expansion. 

For two equal disks, using 
 \eqref{corrShaw} and \eqref{valorecgasint} the force at small distance is 
easily computed:
\be F(Q_1,Q_2,\kappa) = \frac{(Q_1+Q_2)^2}{8 a^2}(\gamma -1 -\log\kappa) -  \frac{(Q_1-Q_2)^2}{2 a^2}.
\label{forzaQris}\ee
The force is  {\em repulsive} and logarithmically divergent at small distances, except for $Q_1 = -Q_2$. This peculiar behavior has been
discussed in \cite{mac2} where is shown that the repulsion comes from the redistribution of charges on the electrodes.
This behavior can make the reader puzzled then we offer a simple explanation. The decomposition \eqref{forza2} is just a trivial 
change of variables, so the point is to show that in a simple configuration the first term gives rise to a logarithmic repulsive force.
The simplest choice is $Q1 = Q_2$. In this case it is obvious that at short distances the density charge on each disk differs from the
usual radial density for an equipotential disk
\be \sigma(r) = \frac{Q/a}{\sqrt{a^2 - r^2}}\label{sigma1disk}\ee
only by terms of order $\kappa$. Computing the force by Coulomb law with \eqref{sigma1disk} gives exactly the $\log\kappa$ term in
\eqref{forzaQris}, with the correct coefficient. Unfortunately, neither this procedure nor the approximate solution given in\cite{mac2}
allow the exact computation of the additive constant $\gamma$, and the author has not been able to derive its analytic form, then
the value is fitted from numerical results. 

In view of subsequent generalizations it is useful to cast the equations \eqref{dec1} in an operatorial form. The equations are defined on the Hilbert space ${\mathbb L}^2$ of square integrable functions on the unit interval. Using the familiar Dirac notation, \eqref{dec1} can be written
\[ (1 + {\cal K})\ket{g} = \ket1\,,\qquad (1 - {\cal K})\ket{f} = \ket1\,. \]
Here $\ket f$ denotes the vector in ${\mathbb L}^2$ represented by the function $f$, $\ket 1$ is the representative of the unit function etc. ${\cal K}$ is the integral operator with kernel $K$.
In this notation the solutions of \eqref{dec1} have the form
\be \ket f \equiv \ket{f_L} = \dfrac1{1-{\cal K}} \ket 1\,,\qquad \ket g = \dfrac1{1+{\cal K}} \ket 1\,,\label{dec1ket1}\ee
and the capacitances are
\be C = \frac a\pi \scalar{1}{f_L}\,,\qquad C_g = \frac {2a}\pi \scalar{1}{g}\,.\label{dec1ket2}\ee
The scalar product $\scalar {f_1}{f_2} $ is the usual one in ${\mathbb L}^2$.
In this form the completely different behavior of $C$ and $C_g$ at small distances is clearly understood.
In this limit ${\cal K} \to 1$ and this produces a divergence in $\ket f$ while $\ket g$ remains bounded. 
There is a corresponding effect in the numerical solution of the equations.
In a numerical computation an approximation, ${\cal K}_0 = {\cal K} + \delta {\cal K}$, is used for the kernel 
and from \eqref{dec1ket1} it is evident that for $\kappa\to 0$ errors are amplified in the equation for $f$, while we expect a much better convergence for $g$, this is confirmed in the numerical solutions.

Before leaving this short review of Love's equation we have to mention a further point, which will be useful below. This equation
can be written in a more compact form
\be F_L(x) - \int_{-1}^1 Q(x,y;\kappa) F_L(y) dy = 1\,;\qquad  Q(x,y;\kappa) = \frac{\kappa}\pi\dfrac1{\kappa^2+ (x-y)^2}\,. \label{complove1.1}\ee
This is due to the symmetry property $K(x,y) = K(y,x) = K(-x,y)$ of the original kernel. $F_L$ is just the extension of $f_L$ to the interval $(-1,1)$ with $f_L(-x) = f_L(x)$.
For a general equation of the type 
\be F(x) - \int_{-1}^1 Q(x,y;\kappa) F(y) dy = h(x) \label{complove1.2}\ee
with $h$ even, it can be shown, using the methods of\cite{kac}, that the solution for $\kappa\to 0$ is given by\cite{Hutson}
\be F(x) = \frac1{\kappa} \int_{-1}^1 {\cal L}(x,y) h(y) dy \label{complove1.3}\ee
where
\be {\cal L}(x,y) = \frac{1}{2\pi} \log\dfrac{1- x y + \sqrt{1-x^2}\sqrt{1-y^2}}{1-x y - \sqrt{1-x^2}\sqrt{1-y^2}}\,. 
\label{complove1.4}\ee
For the same symmetry properties quoted above this solution gives the even extension to the interval $(-1,1)$ of the solution of the equation
\be f(x) - \int_{0}^1 K(x,y) f(y) dy = h(x)\,. \label{complove1.5}\ee
This point will be useful to compute the asymptotic behavior of $C_{ij}$ for different disks.

In the following section  we push the numerical precision to agree with \eqref{corrShaw} in the region of low $\kappa$ and to confirm \eqref{valorecgasint}.

\section{Equal disks: numerical procedure and results\label{eqdisksnum}}

\subsection{Numerical procedure\label{lovenumerical}}
There are several ways to compute numerical solutions of integral equations, we choose one of the most simple methods: a grid of points.
In general an integral can be computed by defining a set of abscissas $x_i$ and corresponding weights $w_i$ and writing, as an instance
\be \int_0^1 F(t) dt \simeq \sum_i F_i w_i \label{integpesi}\ee
where $F_i \equiv F(x_i)$ are the values of the function computed at the abscissas $x_i$, in the interval $(0,1)$. 
With this procedure the equation \eqref{eqaccoppiateug1.3love} on a grid of $N$ points
gives a linear system of $N$ equations:
\be U_i = X_i - \sum_j K_{ij} w_j X_j \label{integpesi2}\ee
where $U_i = 1,\forall i$ and
\be K_{ij} = \frac{\kappa}{\pi}\left( \dfrac{1}{(x_i-x_j)^2 + \kappa^2} + \dfrac{1}{(x_i+x_j)^2 + \kappa^2}\right)\,.
\label{integpesi3}\ee
We used the notation $X_i = f_L(x_i)$ for simplicity.
Between the many possibilities for weights and abscissas we have chosen the method of Gauss's points, and as a check for large $\kappa$ the weights of the trapezoidal rule for integrals.

Once obtained the solution of \eqref{integpesi2} we can compute $C$:
\be C = \frac a\pi \int_0^1 f_L(t) dt \simeq \frac a\pi \sum_i X_i w_i \,.\label{integpesi4}\ee
This method is extremely fast and stable for not too small $\kappa$, let's say $\kappa\geq 0.001$, but 
for small $\kappa$
suffers a slowing down in the convergence as $N$ grows  and, worse, the method becomes unstable. This is expected as the kernel $K(t,s)$ becomes singular in this limit: it is a lorentzian curve which shrinks
to a $\delta$-function. The cure is to regularize\cite{baker,wintle1985edge} the integral in the form
\be \int_0^1 K(t,s) F(s) ds = F(t) \int_0^1 K(t,s)ds + \int_0^1 K(t,s) (F(s) - F(t) )ds \,.\label{metgendi1}\ee
The first integral can be computed analytically, the slightly generalized result is
\be \int_0^\beta K(t,s) ds \equiv G(t;\kappa,\beta) = \frac1\pi \left(\arctan\frac{\beta-t}{\kappa} + \arctan\frac{\beta+t}{\kappa}\right)\,.
\label{valGb}\ee
With this prescription the equation \eqref{integpesi2} now reads
\be U_i = X_i - \sum_j K_{ij} w_j (X_j - X_i) -  G^{(1)}_i X_i\label{integpesi5}\ee
where $G^{(1)}_i \equiv G(x_i;\kappa,1)$. The solution of this equation has no instability for $\kappa\to 0$ but of course the substitution \eqref{metgendi1} cannot cure the slowness in convergence for large $N$ in the regime of small $\kappa$.
In literature it has been verified that one can reach a reasonable stability in the results for $N\,\kappa\gtrsim2$--$3$,
so a direct computation for $\kappa \lesssim 0.0001$ requires a large amount of memory and an extrapolation method is needed to have accurate results. We will use the clever method elaborated  in  the work \cite{Norgren}.
The procedure can be summarized as follows, referring to ref.\cite{Norgren} for more information on the method:
\begin{itemize}
\item[a)] Choose a sequence of decreasing $\kappa$: $\kappa_1, \kappa_2 \ldots$ and a maximum number of grid points, $N_{max}$.
It is supposed that computations can be done for $N<N_{max}$.
Let us call $S_i(N)$ the numerical result obtained  for the capacity for the $i$-th term in the above sequence of $\kappa$'s using a grid of $N$ points.
\item[b)] For each $i$ the best numerical result is $S_i(N_{max})$. Let us note that if the sequence starts with a not too small $\kappa$, e.g. $\kappa = 0.01$, the numerical results reach a stable limit for $N\ll N_{max}$, i.e. these results can be considered as the correct estimate of the true values.

For small $\kappa$ the extrapolated value $C_i$ for the capacitance is given by
\be C_i = S_i( N_{max}) + \left[C_{i-1} - S_{i-1}\left(\frac{\kappa_{i}}{\kappa_{i-1}} N_{max}\right)\right] \label{proceduraN}\ee
This amounts to say that the error is a function of the product $\kappa N$ for $N$ sufficiently large.
We have checked this procedure by comparing sequence of results for different $N_{max}$.
\end{itemize}
In the computation for equal disks we used a maximum number of points $N_{max} = 50000$, this huge number is needed only for $\kappa \leq 0.00005$.

\subsection{Results}
For large distances, $\kappa\gg 1$, the numerical solution of equations \eqref{eqaccoppiateug1.2} converges very fast and gives accurate results, 
so it would be appropriate to take these results as a point of comparison for alternative methods used in the calculation of capacity,
in particular for moment's method which has a wide range of applicability and it is very flexible. The capacitance coefficients can be
expanded to an arbitrary order in powers of $1/\ell$, both for equal and different\cite{maxw2,paf} disks, and the numerical
results are in agreement with the theory. A more detailed analysis shows that the more general expansion in powers of $1/\ell$ valid for arbitrary conductors\cite{mac} describe accurately the data for $\ell$ greater then the diameter of the disks.

In this paper we focus on the more difficult problem of the behavior at small distances of the capacity coefficients.
We performed  calculations down to $\kappa = 5\times10^{-6}$ and the extrapolation procedure outlined above was used for low $\kappa$.
The computed values for $C$ and $C_g$, for a few values of $\kappa$, are given in table~\ref{tabella1V3}, where crude numerical values and extrapolated values are reported and compared with the theoretical calculations.
We verified that the adopted numerical procedure always satisfy the lower bound given by Ignatovsky approximation.
\begin{table}[!ht]
{\small
\noindent
\begin{tabular}{llllllll}
   $\kappa$ & $C$ (num) & $C$ (extrap.) & $\;\;C_K + C_K^{(1)} $  & $\qquad C_K$ & $\qquad C_g$ \\[5pt]
   0.1 & 2.93898079847 & 2.93898079847 & 2.93861918035 & 2.91538673825 & 0.34115344081 \\
   0.09 & 3.22351516784 & 3.22351516784 & 3.22321135328 & 3.20154883946 & 0.33932828637 \\
   0.08 & 3.57841124252 & 3.57841124252 & 3.57816138551 & 3.55814393785 & 0.33745070948 \\
   0.07 & 4.03368449989 & 4.03368449989 & 4.03348452192 & 4.01519859988 & 0.33551381399 \\
   0.05 & 5.48515774661 & 5.48515774661 & 5.48504415268 & 5.47054563828 & 0.33142337201 \\
   0.04 & 6.75077356422 & 6.75077356422 & 6.75069575380 & 6.73830283789 & 0.32924098908 \\
   0.03 & 8.85466722892 & 8.85466722892 & 8.85461962224 & 8.84452918316 & 0.32693591075 \\
   0.02 & 13.0509956667 & 13.0509956667 & 13.0509719921 & 13.0434617379 & 0.32446447909 \\
   0.01 & 25.6031005825 & 25.6031005825 & 25.6030935132 & 25.5986206380 & 0.32173444759 \\
   0.005 & 50.6564073534 & 50.6564073534 & 50.6564052748 & 50.6537795380 & 0.32019662884 \\
   0.002 & 125.727970864 & 125.727970864 & 125.727970460 & 125.726695638 & 0.31915711744 \\
   0.001 & 250.782584053 & 250.782584053 & 250.782583938 & 250.781854538 & 0.31876855936 \\
   0.0005 & 500.837427207 & 500.837427207 & 500.837427175 & 500.837013438 & 0.31855676457 \\
   0.0002 & 1250.91012284 & 1250.91012284 & 1250.91012283 & 1250.90992954 & 0.31841791733 \\
   0.0001 & 2500.96519631 & 2500.96519631 & 2500.96519630 & 2500.96508844 & 0.31836741251 \\
   0.00005 & 5001.02030939 & 5001.02030719 & 5001.02030718 & 5001.02024734 & 0.31834040490 \\
   0.00003 & 8334.39443515 & 8334.39426957 & 8334.39426952 & 8334.39423088 & 0.31832897372 \\
   0.00002 & 12501.0947926 & 12501.0931909 & 12501.0931907 & 12501.0931634 & 0.31832302200 \\
   0.000015 & 16667.7880979 & 16667.7827448 & 16667.7827444 & 16667.7827231 & 0.31831995668 \\
   0.00001 & 25001.1684013 & 25001.1483383 & 25001.1483373 & 25001.1483223 & 0.31831680531 \\
   0.000005 & 50001.3060052 & 50001.2034999 & 50001.2034894 &
   50001.2034812 & 0.31831352139 \\
\end{tabular}
}
\caption{Numerical ($N=50000$) and extrapolated  values of $C/a$. $C_K$ is the leading Kirchhoff formula. For $C_g$ the first 11 digits are identical for the numerical and the extrapolated versions.\label{tabella1V3}}
\end{table}

\begin{figure}[ht]
\begin{center}
\includegraphics[width=0.495\textwidth]{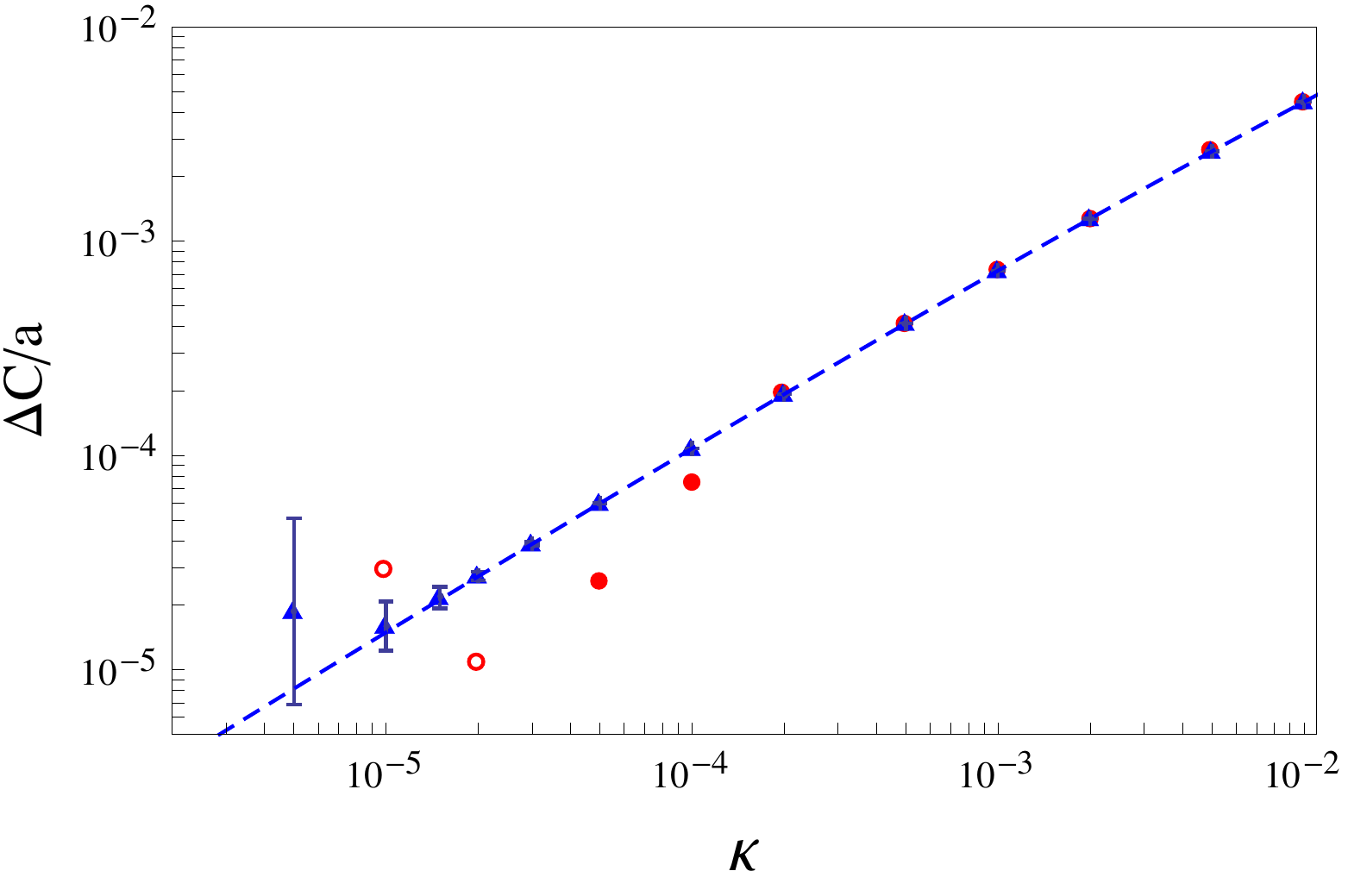}
\includegraphics[width=0.495\textwidth]{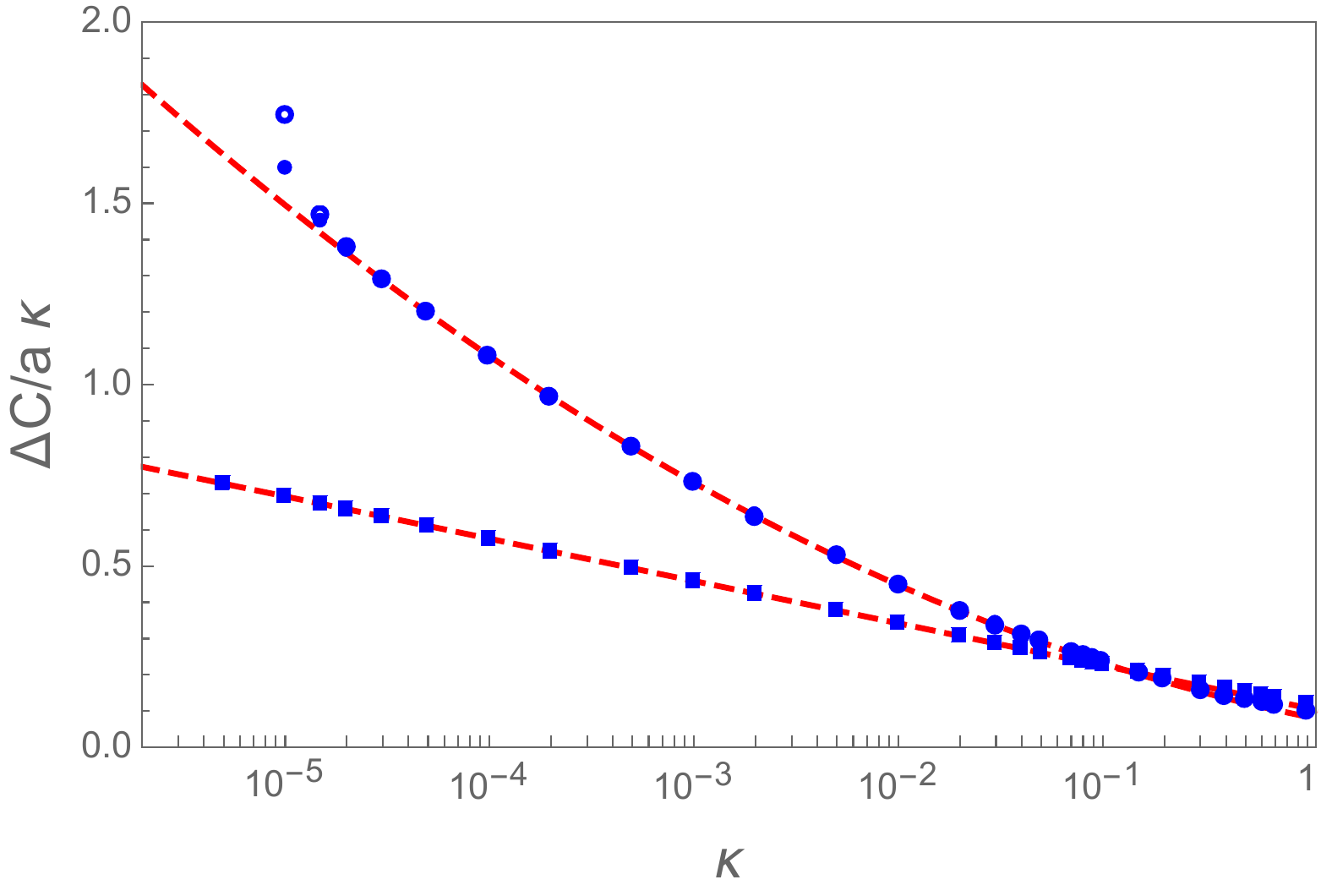}
\caption{Left panel: Difference $(C - C_K)/a$ (triangles).
Disks and empty disks are the points of ref.\cite{Norgren}, positive and negative respectively.
Dashed line is the theoretical next order correction in \eqref{corrShaw}. Right panel:
$(C - C_K)/a\kappa$ (points) and $(C_{g}/a-1/\pi)/\kappa$ (squares). $N=50\cdot10^3$ (filled disks) and $N=45\cdot10^3$ (empty disks).
The dashed line is the theoretical next order corrections in \eqref{corrShaw} and \eqref{valorecgasint}.
\label{figura1}}
\end{center}
\end{figure}

A graphical representation of the results is given in figure~\ref{figura1}.
In the left panel it is shown the difference between the numerical results for $C$ and the Kirchhoff approximation $C_K$.
The dashed line is the expected result $C_K^{(1)}$, see equation \eqref{corrShaw}. In the same figure we reported the results
of the work \cite{Norgren}, obtained by another method and with a smaller grid (15000 points): the two sets agree up to $\kappa\sim 10^{-4}$, for smaller distances the data of ref.\cite{Norgren} start to loss precision, probably due the too small grid.

We compare now our numerical values to theoretical expectations performing a rather severe test: we plot the the ratios $(C - C_K)/a\kappa$ and 
$(C_g - a/\pi)/a\kappa$ versus $\kappa$. In a logarithmic scale  a parabola for the first quantity and a straight line for the second
must survive.
The results are shown in the right panel of figure~\ref{figura1}. 
The agreement covers several orders of magnitude, and it is clear that the division by $\kappa$ amplifies every possible error. The effect is visible
on the lowest $\kappa$ value for $C$: while from table~\ref{tabella1V3} the agreement appears excellent, from the figure it is clear that the point is slightly overestimated. 
This is surely due to the limitation in $N$: we have verified that a smaller $N$ tend to overestimate $C$. In figure~\ref{figura1} we show for comparison the results for $N_{max} = 45000$, they are indistinguishable from the higher precision points up to  $\kappa\sim3\times10^{-5}$, for lower $\kappa$ they 
overestimate the results.
The value of $\gamma$ in \eqref{valorecgasint} has been obtained by fitting the values of $C_{g}$ in figure~\ref{figura1}.
Only $\gamma$ is fitted, the slope of the straight line is fixed by \eqref{valorecgasint}.

In conclusion our numerical computation strongly supports the asymptotic estimates \eqref{corrShaw} and \eqref{valorecgasint}, showing  that
these are a good approximation for capacity also for $\kappa$ as high as $\kappa\sim 1$. On the other hand the agreement testifies to the accuracy of the numerical and extrapolation procedures,  on this basis we are confident that the same procedures could be applied to other systems, like capacitors with different disks.

\section{Different disks: theoretical results\label{diffdisksth}}
A generalization of the system described in section~\ref{sezdischiuguali} consists of two disks of radii $a, c$, coaxial, at distance $\ell$. 
In \cite{paf} it has been shown that the Dirichlet problem for this system can be reduced to the solution of a system of integral equations, very similar to \eqref{eqaccoppiateug1}
\be
V_1 = F_1(t) +\int_0^\beta K(t,z;\kappa) F_2(z)\,dz\,;\quad 
V_2 = F_2(t) +  \int_0^1 K(t,z;\kappa) F_1(z)\,dz\,.
\label{eqaccoppiatediv1}
\ee
where $\beta = c/a>1$ and $\kappa = \ell/a$. The kernel $K$ is the same as before, see eq.\eqref{kernelsing}. The charges on the disks
are given by
\be Q_1 = \frac{2a}{\pi} \int_0^1 F_1(t)\,dt\,;\quad Q_2 = \frac{2a}{\pi} \int_0^\beta F_2(t)\,dt\,.\label{caichedef}\ee
The functions $F_i$ are related to charge densities by an Abel transformation\cite{sne,Carlson,paf}
\begin{align}
&F_1(t) = 2\pi a \int_t^1\dfrac{x}{\sqrt{x^2-t^2}}\,\sigma_1(x)\,dx\,;\quad F_2(t) = 2\pi a \int_t^\beta\dfrac{x}{\sqrt{x^2-t^2}}\,\sigma_2(x)\,dx\,;\label{abeltransf}\\
&\sigma_1(x) = \frac1{a\pi^2}\left[ \dfrac{F_1(1)}{\sqrt{1-x^2}} - \int_x^1 dt \dfrac{F_1'(t)}{\sqrt{t^2-x^2}}\right]\,;
\quad
\sigma_2(x) = \frac1{a\pi^2}\left[ \dfrac{F_2(\beta)}{\sqrt{\beta^2-x^2}} - \int_x^\beta dt \dfrac{F_2'(t)}{\sqrt{t^2-x^2}}\right]\,.
\nonumber
\end{align}
From \eqref{eqaccoppiatediv1} and \eqref{caichedef} it follows that by solving the two systems
\begin{subequations}\label{sistemifg}
\begin{align}
&1 = f_1(t) + \int_0^\beta K(t,s;\kappa) f_2(s)\,ds \,;\qquad 0 = f_2(t) + \int_0^1 K(t,s;\kappa) f_1(s)\,ds\label{sistemifg1a}\\
&0 = g_1(t) + \int_0^\beta K(t,s;\kappa) g_2(s)\,ds \,;\qquad 1 = g_2(t) + \int_0^1 K(t,s;\kappa) g_1(s)\,ds\label{sistemifg1b}
\end{align}
\end{subequations}
one can compute the capacitance coefficients by integrating the solutions:
\be \begin{split}
&C_{11} = \frac{2a}{\pi}\int_0^1 f_1(t) dt\,;\qquad C_{21} = \frac{2a}{\pi}\int_0^\beta f_2(t) dt\,;\\
&C_{12} = \frac{2a}{\pi}\int_0^1 g_1(t) dt\,;\qquad C_{22} = \frac{2a}{\pi}\int_0^\beta g_2(t) dt\,.
\end{split}
\label{capdiverse}
\ee
In some applications it can be useful to write the equations fixing the charges, instead of the potentials.
A simple transformation\cite{mac2} of \eqref{eqaccoppiatediv1} gives
\begin{equation}
\begin{split}
\frac{\pi Q_1}{2a}  &= F_1(t) + \int_0^\beta K(t,s)F_2(s)ds- \int_0^\beta G(s;\kappa,1)) F_2(s)ds\\
\frac{\pi Q_2}{2a \beta}  &= F_2(t) + \int_0^1 K(t,s)F_1(s)ds -\frac1\beta\int_0^1 G(s;\kappa,\beta) F_1(s) ds\end{split}\label{potcof2div}
\end{equation}
The function $G$ has been defined in \eqref{valGb}.

In \cite{paf} the leading order in the short distance expansion for $C_{ij}$ has been computed:
\begin{subequations}\label{valoriCij}
\begin{align}
&C_{11}^{(0)} = \left\{\frac{a}{4\kappa} + \frac{a}{2\pi}\left[ \log\frac{8\pi}\kappa - 1\right]\right\}
+ \frac{a}{\pi}\left[ \beta - \sqrt{\beta^2-1} - \frac12 {\rm arctanh}\frac1\beta\right]\\
&C_{12}^{(0)} = -\left\{\frac{a}{4\kappa} + \frac{a}{2\pi}\left[ \log\frac{8\pi}\kappa - 1\right]\right\}
+ \frac{a}{2\pi} {\rm arctanh}\frac1\beta\\
&C_{22}^{(0)} = \left\{\frac{a}{4\kappa} + \frac{a}{2\pi}\left[ \log\frac{8\pi}\kappa - 1\right]\right\}
+ \frac{a}{\pi}\left[ \beta + \sqrt{\beta^2-1} - \frac12 {\rm arctanh}\frac1\beta\right]
\end{align}
\end{subequations}
The corrections to \eqref{valoriCij} are $o(1)$, i.e. these expressions are the asymptotic forms of capacitance coefficients for $\kappa\to 0$. The expression in curly brackets is the double of the leading order for the capacitance of two equal disks at distance $2\ell$, as explained in \cite{paf}.
The reader has probably noticed that in \eqref{valoriCij}
the logarithmic corrections to the geometric capacitance (the term in $1/\kappa$) are different from the Kirchhoff approximation for equal disks, 
then for $\beta\to 1$ one {\em does not} get \eqref{corrShaw}. From the mathematical point of view this is not a problem: the limit $\kappa\to 0$ is defined with $\beta>1$ fixed, so simply the limits do not commute. From a physical point of view the question as some importance as $\beta=1$, exactly, is a mathematical fiction: this means that a crossover region must exists when $\beta\sim 1$, where the approximation \eqref{valoriCij} becomes
valid only for very small $\kappa$. This crossover region will be investigated below.

Following the general philosophy outlined in the introduction we look for combinations of $C_{ij}$ which can be distinguished by their different behavior for $\kappa\to 0$.

It is tempting to generalize to this system the same variables used \eqref{defc}. 
In\cite{paf}
it is argued that, in the general case, the combinations $C_{11}+ C_{12}$ and $C_{12}+C_{22}$ are separately finite for $\kappa\to 0$ while, 
analogously to \eqref{vallimcg} their sum tends to $C_T$, the self capacitance of the collapsed system. Then we will explore the variables
\be C_{g1} = C_{11}+ C_{12}\,;\qquad C_{g2} = C_{22}+ C_{12}\,; \qquad C_g =  C_{g1} + C_{g2}\,.
\label{vallimcg2}\ee
For a two disks capacitor $C_T$ is the capacity of the larger disk, i.e. $C_T = 2 c/\pi\equiv 2 a \beta/\pi$. 
The smooth behavior of the quantities \eqref{vallimcg2} is verified by the explicit computation in the  particular system under study, from \eqref{valoriCij} for $\kappa\to 0$ we have in effect:
\be C_{g1} = \frac{a}{\pi}(\beta - \sqrt{\beta^2-1})\,+ o(1) ;\quad C_{g2} = \frac{a}{\pi}(\beta + \sqrt{\beta^2-1})\,+ o(1)\,;
\quad C_g = \frac{2 a \beta}{\pi} + o(1)\,.
\label{limitcg1cg2}
\ee
In this work we compute the next order in $\kappa$ to equations \eqref{valoriCij}.
 This is a relevant point for forces. In \cite{mac2} it has been pointed out that the force between two conductors can be written in the form
\be
F = \frac12 \dfrac{(Q_a+ Q_b)^2}{(C_{g1}+ C_{g2})^2} \dfrac{\partial}{\partial\ell}(C_{g1}+ C_{g2})
- \frac12 \dfrac{\partial}{\partial\ell}\left[\dfrac{ (C_{g2} Q_a - C_{g1} Q_b)^2}{(C_{g1}+ C_{g2})^2}\frac1{C} \right]
\label{forzagen}
\ee
As $C$ diverge at short distances as $1/\ell$ (for plates) the first term in \eqref{forzagen} 
can be relevant,
except for the particular case $Q_b=-Q_a$,
and this term depends on the corrections to \eqref{limitcg1cg2}.

The final results for the capacitance coefficients are
\be C_g = C_{g1}+ C_{g2} =
\frac{2 a \beta}{\pi} + a B\, \kappa
\label{risfin1}
\ee
\be
\begin{split}
&C_{g1}=
C_{11} + C_{21} = a\Bigl\{
\frac1{\pi} \left( \beta - \sqrt{\beta^2-1}\right) + \frac{\kappa}{\pi^2\sqrt{\beta^2-1}}\log \frac{8\pi}{\kappa} + \kappa X_{g1}\Bigr\}\\
&C_{g2}=C_{12} + C_{22} = a\Bigl\{
\frac1{\pi} \left( \beta + \sqrt{\beta^2-1}\right) - \frac{\kappa}{\pi^2\sqrt{\beta^2-1}}\log\frac{8\pi}{\kappa} + \kappa(B - X_{g1})\Bigr\}
\end{split}
\label{risfin2}\ee
and
\begin{multline}
C_{11} = 2 C_{EQ}(2\kappa) + 
a \left\{\frac1{\pi} \left(\beta - \sqrt{\beta^2-1}- \frac12\arctanh\frac{1}{\beta}\right)
\right.\\
\left.
- \frac{\kappa}{2\pi^2} \left(\frac{\beta}{\beta^2-1}-\frac{2}{\sqrt{\beta^2-1}}+\arctanh\frac{1}{\beta}\right) 
 \log \frac{8 \pi }{\kappa}
   +
   \kappa\,( Y_1 +  Y_2)\right\}
   \label{valoreC11def}
\end{multline}
Where $2 C_{EQ}(2\kappa)$ is the double of the relative capacitance of two equal disks at distance $2\ell$, the direct generalization of the first term in \eqref{valoriCij}. From \eqref{corrShaw}:
\be 2 C_{EQ}(2\kappa) = 
a\left\{ \frac1{4\kappa} + \frac1{2\pi}\left[\log\frac{8\pi}{\kappa} - 1\right]
+\frac1{4\pi^2} \kappa\left[ \left(\log\frac{\kappa}{8\pi}\right)^2-2\right]
\right\}
\label{valoreceq2k}
\ee
The proof of these results 
and the explicit value for the constants $B, X_{g1}, Y_{1}, Y_{2}$
are given in section \ref{sezioneris}. 

With $C_{11}$ at our disposal we can calculate each coefficient $C_{ij}$ up to the first order in $\kappa$:
\be
\begin{split}
& C_{12} = C_{g1} - C_{11}\,;\qquad C_{22} = C_{g2} - C_{g1} + C_{11}\,;\\
 &C = \dfrac{C_{11} C_{22} - C_{12}^2}{C_{11} + C_{22} + 2C_{12}}
= \dfrac{C_{g1} C_{g2}}{C_g} - C_{12} = \dfrac{C_{g1} C_{g2}}{C_g} - C_{g1} + C_{11}\,.
\end{split}
\label{valoreCij2pert}
\ee
From these results it follows the predicted hierarchy in the short distance bahvior: $C_g$ has no logarithmic corrections, $C_{g1}$ and $C_{g2}$ have a leading correction of the form $\kappa\log\kappa$ and $C_{11}$ a leading correction of the form $\kappa\log^2\kappa$.

The theoretical and numerical computations are greatly simplified by decoupling the systems \eqref{sistemifg}, 
in analogy with the procedure adopted for equal disks.
For the couple $f_1, f_2$
substitution of the second equation in the first and a manipulation of the integrals gives for  $f_1$:
\be
1 = f_1(t) - \int_0^1 K(t,x;2\kappa) f_1(x) dx + \int_\beta^\infty K(t,s;\kappa)ds \int_0^1 dx K(s,x;\kappa) f_1(x) \,.
 \label{appb1.3}\ee
 This is a linear integral equation for $f_1$ expressed in terms of the two kernels
 \be A(t,x) = K(t,x;2\kappa)\,;\qquad B(t,x) = \int_\beta^\infty K(t,s;\kappa) K(s,x;\kappa)\,ds
 \label{twokernels}\ee
 and can be numerically solved by usual techniques.
 A similar transformation can be done for the couple $g_1, g_2$, obtaining 
\be - G(t;\kappa,\beta) = 
g_1(t) - \int_0^1 K(t,x;2\kappa) g_1(x) dx + \int_\beta^\infty K(t,s;\kappa)ds \int_0^1 dx K(s,x;\kappa) g_1(x) \,.
 \label{appb1.3g1}\ee
Once obtained $f_1, g_1$, one can compute $f_2, g_2$ from equations~\eqref{sistemifg} and  all coefficients $C_{ij}$ can be obtained by integration.

The integral defining the kernel $B$ in \eqref{twokernels} can be analytically computed and gives:
\be
\begin{split}
B(t,x) &= \frac{\kappa}{\pi}\left(\frac{1}{4 \kappa^2+(t+x)^2}+\frac{1}{4 \kappa^2+(t-x)^2}\right)\left(2 - G(t;\kappa,\beta) - G(x;\kappa,\beta)\right)\\
&+ \frac{\kappa^2}{\pi^2} \Bigl(\dfrac{ \log \left(\frac{(\beta-t)^2+\kappa^2}{(\beta-x)^2+\kappa^2}\right)}{(t-x)
   \left(4 \kappa^2+(t-x)^2\right)}+\frac{ \log
   \left(\frac{(\beta-x)^2+\kappa^2}{(\beta+t)^2+\kappa^2}\right)}{(t+x) \left(4
   \kappa^2+(t+x)^2\right)}\\
   &+\frac{ \log
   \left(\frac{(\beta-t)^2+\kappa^2}{(\beta+x)^2+\kappa^2}\right)}{(t+x) \left(4
   \kappa^2+(t+x)^2\right)}+\frac{ \log
   \left(\frac{(\beta+x)^2+\kappa^2}{(\beta+t)^2+\kappa^2}\right)}{(t-x) \left(4
   \kappa^2+(t-x)^2\right)}\Bigr)
\end{split}
\label{valoreB}
\ee
There is not a singularity for $x\to t$, performing the limit:
\begin{multline} B(t,t) = \frac1{2\kappa\pi^2}
\left( \frac{\kappa (t-\beta)}{(\beta-t)^2+\kappa^2}-\frac{\kappa (\beta+t)}{(\beta+t)^2+\kappa^2}+\frac{\pi  \kappa^2}{\kappa^2+t^2}+\pi \right) +\\
\frac{1}{8\kappa\pi^2} \frac{2 \kappa^3 \log \left(1-\frac{4 \beta t}{(\beta+t)^2+\kappa^2}\right)-4 t \left(2 \kappa^2+t^2\right)
   \left(\arctan\frac{\beta-t}{\kappa}
   +\arctan\frac{\beta+t}{\kappa}\right)}{t \left(\kappa^2+t^2\right)}
   \label{btt}
\end{multline}
The singularity for $t=0$ also cancels, as the reader can easily verify.
\section{Different disks: numerical procedure and results\label{diffdisksnum}}
The guidelines for the calculations in the case of different disks are the same as those set out in section~\ref{eqdisksnum}.
The computation is based on the decoupled equations \eqref{appb1.3} and \eqref{appb1.3g1}, the system \eqref{sistemifg} has been used to check the results.

We first compare the numerical calculations of $C_{ij}$ with the theoretical predictions
\eqref{valoriCij} at small distances. A set of representative values for $C_{ij}$ are given in table~\ref{tabc11c12c22}.
\noindent
\begin{table}[!ht]
\begin{tabular}{crrrr}
  &  &  &  &  \\
 $\kappa$ & 0.01 & 0.005 & 0.002 & 0.001 \\
$C_{11}$[num]  & 26.0577642616 & 51.1647446660 & 126.307910662 & 251.417074438 \\
$C_{11}$  & 26.0577642616 & 51.1647446660 & 126.307910662 & 251.417074438 \\
\eqref{valoreC11def},\eqref{valoreCij2pert}  & 26.0582443232 & 51.1649152968 & 126.307952936 & 251.417089266 \\ \hline \\[-7pt]
$C_{22}$[num]  & 26.3272784774 & 51.4438960868 & 126.593829377 & 251.705602453 \\
$C_{22}$  & 26.3272784774 & 51.4438960868 & 126.593829377 & 251.705602453 \\
\eqref{valoreC11def},\eqref{valoreCij2pert}  & 26.3282180567 & 51.4442375233 & 126.593912954 & 251.705630678 \\ \hline \\[-7pt]
$C_{12}$[num]  & -25.8414180520 & -50.9537146921 & -126.100547790 & -251.211107771 \\
$C_{12}$  & -25.8414180520 & -50.9537146921 & -126.100547790 & -251.211107771 \\
\eqref{valoreC11def},\eqref{valoreCij2pert}  & -25.8422031124 & -50.9539919338 & -126.100614630 & -251.211130377 \\
\end{tabular}

\begin{tabular}{crrrr}
  &  &  &  &  \\
$\kappa$ & 0.0005 & 0.0002 & 0.0001 & 0.00005 \\
$C_{11}$[num]  & 501.526693409 & 1251.67201670 & 2501.78212974 & 5001.89232927 \\
 $C_{11}$ & 501.526693409 & 1251.67201670 & 2501.78212974 & 5001.89232927 \\
\eqref{valoreC11def},\eqref{valoreCij2pert}  & 501.526698761 & 1251.67201820 & 2501.78213035 & 5001.89232953 \\ \hline \\[-7pt]
$C_{22}$[num]  & 501.816674562 & 1251.96297043 & 2502.07344404 & 5002.18383906 \\
$C_{22}$  & 501.816674562 & 1251.96297043 & 2502.07344404 & 5002.18383906 \\
\eqref{valoreC11def},\eqref{valoreCij2pert}  & 501.816684127 & 1251.96297280 & 2502.07344492 & 5002.18383940 \\ \hline \\[-7pt]
$C_{12}$[num]  & -501.321498455 & -1251.46733489 & -2501.57763713 & -5001.68793885 \\
$C_{12}$  & -501.321498455 & -1251.46733489 & -2501.57763713 & -5001.68793885 \\
  & -501.321506209 & -1251.46733688 & -2501.57763789 & -5001.68793915 \\
\end{tabular}

\begin{tabular}{crrrr}
 &  &  &  &  \\
 $\kappa$ & 0.00003 & 0.00002 & 0.000015 & 0.00001 \\
 $C_{11}$[num] & 8335.30690980 & 12502.0380948 & 16668.7506873 & 25002.1503769 \\
 $C_{11}$ & 8335.30690965 & 12502.0380790 & 16668.7505162 & 25002.1483652 \\
\eqref{valoreC11def},\eqref{valoreCij2pert}  & 8335.30690978 & 12502.0380791 & 16668.7505162 & 25002.1483649 \\ \hline \\[-7pt]
$C_{22}$[num]  & 8335.59850325 & 12502.3297322 & 16669.0423473 & 25002.4420604 \\
$C_{22}$  & 8335.59850309 & 12502.3297163 & 16669.0421763 & 25002.4400488 \\
\eqref{valoreC11def},\eqref{valoreCij2pert}  & 8335.59850326 & 12502.3297164 & 16669.0421763 & 25002.4400484 \\ \hline \\[-7pt]
$C_{12}$[num]  & -8335.10256298 & -12501.8337708 & -16668.5463751 & -25001.9460768 \\
$C_{12}$  & -8335.10256283 & -12501.8337550 & -16668.5462040 & -25001.9440652 \\
 \eqref{valoreC11def},\eqref{valoreCij2pert} & -8335.10256298 & -12501.8337551 & -16668.5462040 & -25001.9440649 \\
\end{tabular}
\caption{Values, with $a=1$ and $\beta=1.1$ of $C_{11}, C_{22}, C_{12}$. We give the rough numerical data for $N=55000$, the extrapolated
values and the asymptotic estimates \eqref{valoreC11def} and \eqref{valoreCij2pert}.\label{tabc11c12c22}}
\end{table}

\begin{figure}[!ht]
\begin{center}
\includegraphics[width=0.70\textwidth]{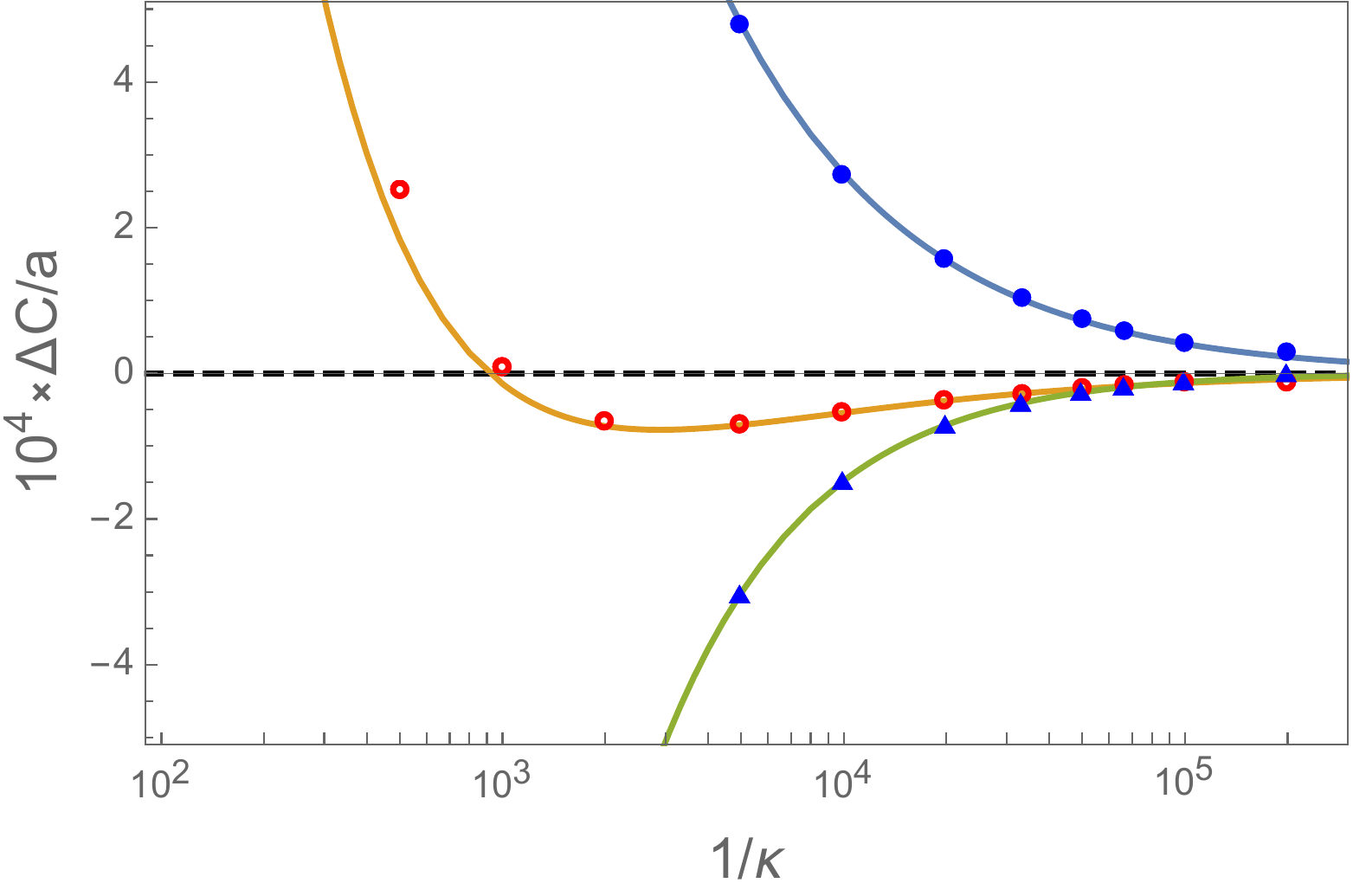}
\caption{Difference between computed and predicted asymptotic values for $\beta=1.1$ and $N_{max} = 55000$. $\Delta C_{11}/a$ (filled disks), 
$\Delta C_{12}/a$ (empty disks), $\Delta C_{22}/a$ (triangles). 
The last point, with $\kappa=5\times10^{-6}$ can have a small error due to extrapolation. The continuous lines are the computed asymptotic values for $(C_{ij} - C^{(0)}_{ij})/a$. The scale is expanded by a factor $10^4$.
\label{dcb1p1S40}}
\end{center}
\end{figure}

In figure~\ref{dcb1p1S40} the difference between the two values amplified by a factor $10^4$ is plotted
against $1/\kappa$ for $\beta=1.1$. First of all the data show that $C_{ij}\to C^{(0)}_{ij}$, confirming the asymptotic analysis of \cite{paf}.
The figure also shows that the approach to the asymptotic values is in agreement with the theoretical predictions 
\eqref{valoreC11def} and \eqref{valoreCij2pert}.  
For other values of $\beta$ the qualitative behavior is similar but the agreement with \eqref{valoriCij} shifts toward smaller values of $\kappa$ for $\beta\to 1$. 

\begin{figure}[!ht]
\begin{center}
\includegraphics[width=0.495\textwidth]{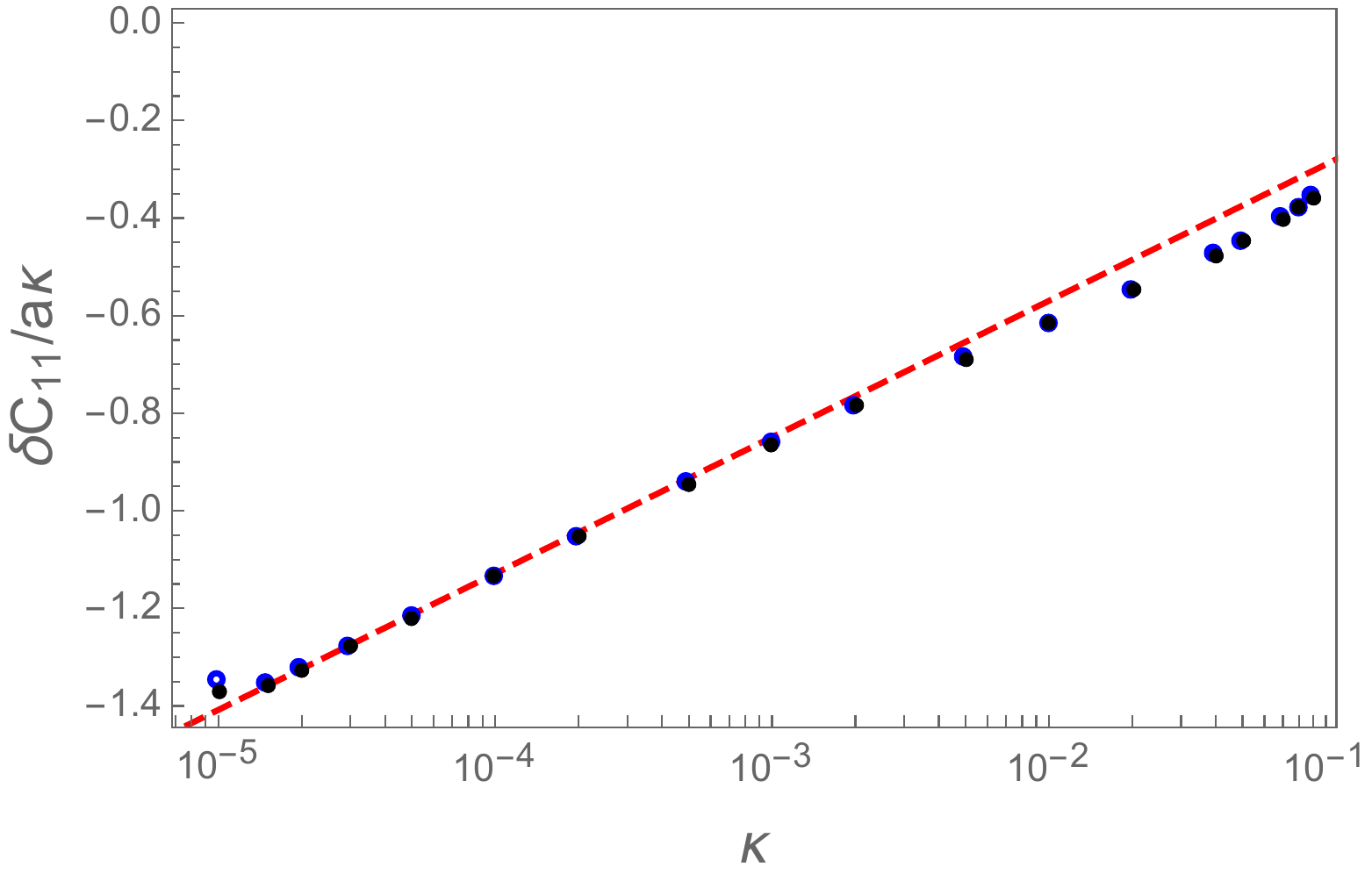}
\includegraphics[width=0.495\textwidth]{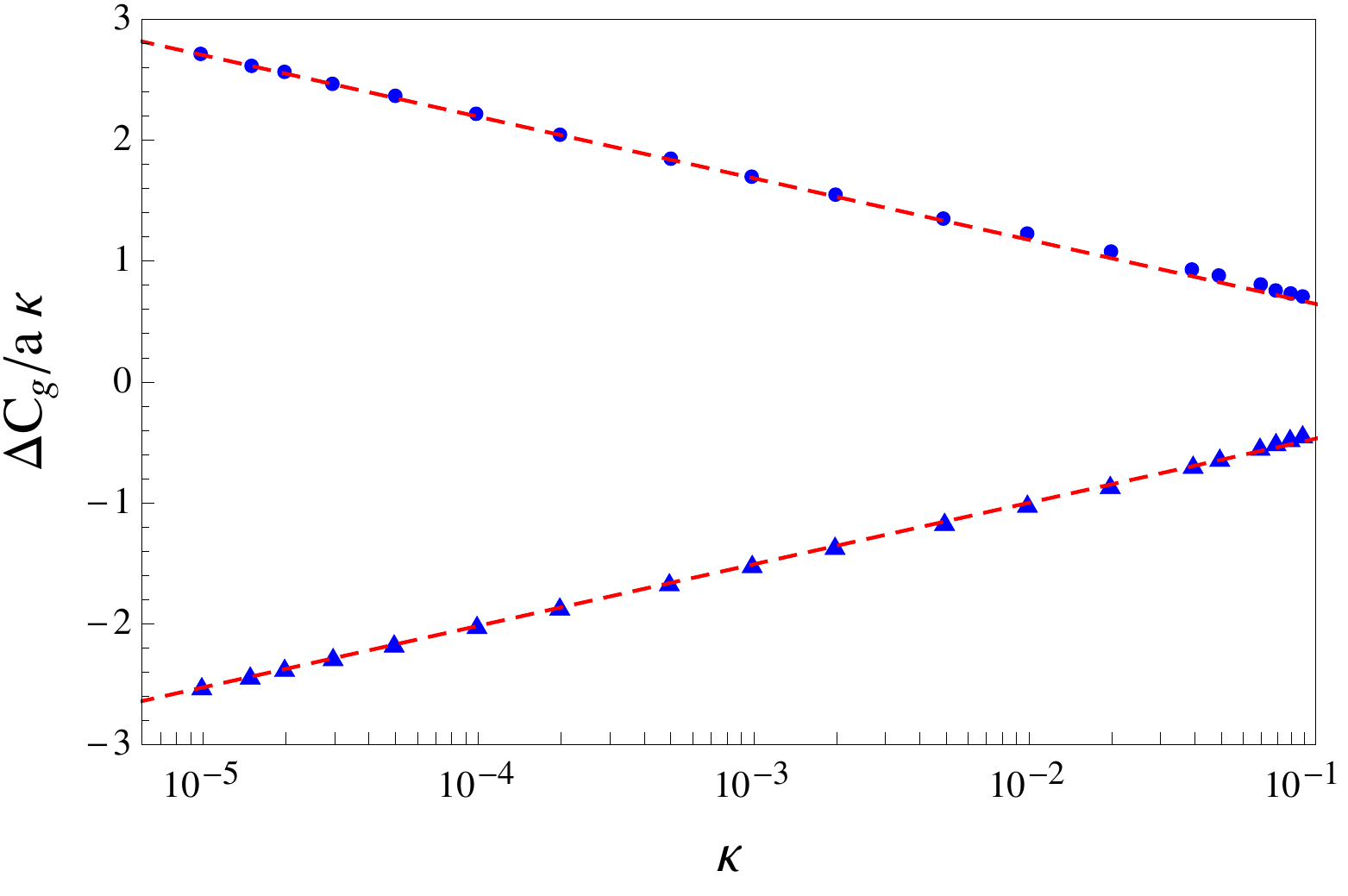}
\caption{Left panel: the quantity $\delta C_{11}/a$, \eqref{deltaC11suk}, vs the perturbation calculation \eqref{pertdeltaC11suk} (dashed line)
for $N=55000$ (filled disks) and $N=45000$ (empty disks).
Right panel: $C_{g1}(\kappa)- C_{g1}(0))/\kappa$ (circles) and $C_{g2}(\kappa)- C_{g2}(0))/\kappa$ (triangles) for $\beta=1.1$.
The dashed curves are the predictions \eqref{valoreasyntcg1} and \eqref{valoreasyntcg2}.
\label{figdeltaC11suk}}
\end{center}
\end{figure}

To perform a direct check of perturbative calculation we consider
\be \frac1{a\kappa}\delta C_{11}=\frac{1}{a\kappa}
\left\{
C_{11}- 2 C_{EQ}(2\kappa) - a\frac1{\pi} \left(\beta - \sqrt{\beta^2-1}- \frac12\arctanh\frac{1}{\beta}\right) \right\}\label{deltaC11suk} \ee
which according to \eqref{valoreC11def} must be given by
\be \frac1{a\kappa}\delta C_{11}=
- \frac{1}{2\pi^2} \left(\frac{\beta}{\beta^2-1}-\frac{2}{\sqrt{\beta^2-1}}+\arctanh\frac{1}{\beta}\right) 
 \log \frac{8 \pi }{\kappa} 
   +\, Y_1(\beta) +  Y_2(\beta)
   \label{pertdeltaC11suk}
   \ee
   i.e. a straight line in a $\log\kappa$ scale. Every possible discrepancy is enhanced by the prefactor $1/\kappa$ in \eqref{deltaC11suk}.
The results are shown in the left panel of figure~\ref{figdeltaC11suk}. 
In our opinion the numerical data support the theoretical analysis: the numerical computations at $N=45000$ and $N=55000$ are indistinguishable down to $\kappa = 10^{-5}$ and follow the theoretical results. The first point is numerically slightly overestimated 
even at these large values of $N$, as it is clear comparing the results ad the two values of $N$ reported in figure. In absolute
value this residual discrepancy, due to numerical approximations, amounts to three parts in $10^{12}$, as can be verified in table~\ref{tabc11c12c22}.

We can deepen the analysis considering the combinations $C_{g1}, C_{g2}$. According to the theoretical results 
\eqref{risfin2}
these quantities have a finite limit for $\kappa\to 0$ and the approach to the limit is of the form $\kappa(a \log\kappa + b)$, i.e. on a logarithmic scale $(C_{gi}- C_{gi}(0))/\kappa$ must lie on a straight line, with the coefficients $a,b$ fixed in \eqref{risfin2}. The data and the predictions are shown in the right panel of figure~\ref{figdeltaC11suk} and we think that the agreement is excellent:
the test is particularly severe as the division by $\kappa$ enhance any error at small distances by a huge factor.
These results confirm our general expectation: in the combinations $C_{g1}, C_{g2}$ the asymptotic behavior is smoothed.

\begin{figure}[!ht]
\begin{center}
\includegraphics[width=0.75\textwidth]{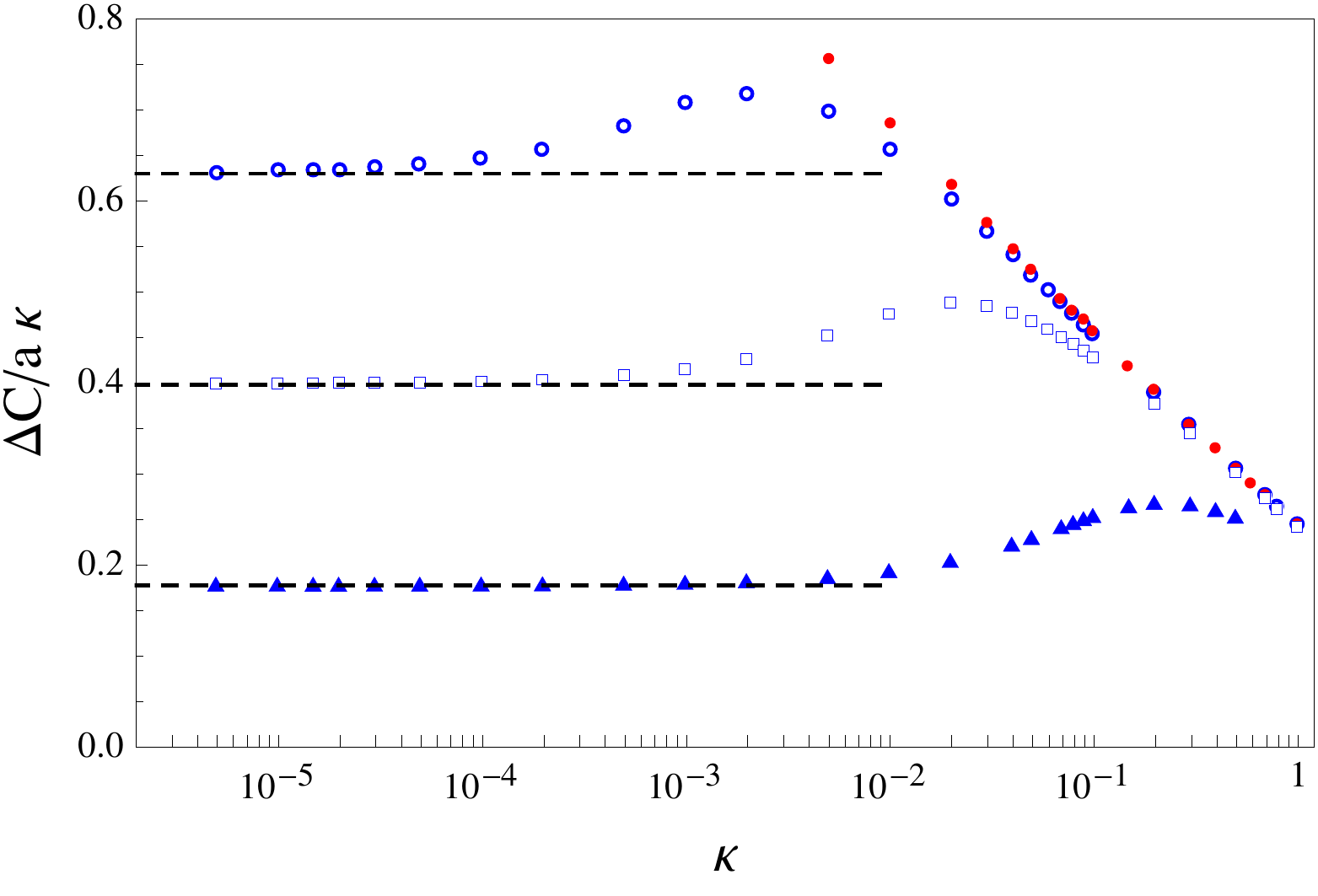}
\caption{The quantity $(C_{g1}+C_{g2} - C_T)/\kappa$ for different values of $\beta$, $\beta=1.1$, triangles,
$\beta=1.01$, empty squares, $\beta=1.001$, empty circles. The filled circles lying on a straight line are the results for
$2(C_{11}+ C_{12})$ for equal disks, i.e. $C_g$ for
 $\beta=1$.
Dashed lines are the theoretical asymptotic values \eqref{valoreBcoeff}.
\label{figura2LogV3}}
\end{center}
\end{figure}

The last step is the consideration of $C_{g} = C_{g1} + C_{g2}$ for which  we expect that at order $\kappa$ even the terms $\kappa\log\kappa$ cancel
out, leaving the result \eqref{sommacgdiff}. We consider again the difference between the numerical value and the asymptotic value divided by $\kappa$: $(C_g - 2a\beta/\pi)/(a \kappa)$  and plot the results in figure~\ref{figura2LogV3} for several values of $\beta$, $\beta=1.1, 1.01, 1.001$ from the bottom to the top.

The agreement between theory and numerical results is again satisfactory, but the numerical analysis also shows rather vividly the crossover mechanism as $\beta\to 1$. 
The coefficient $B$ grows as $\beta\to 1$ and the curves have as an envelope the result for $\beta=1$ (equal disks).
The logarithmic singularity in the case of equal disks is traded for a growing plateau, diverging for $\beta\to 1$. 
This is in agreement with the theoretical calculations. For $\beta\to 1$
\be B \to -\frac{1}{\pi^2}\log(\beta-1) + \text{const.} \label{valoreB1bis}\ee
The analogous of the sum \eqref{sommacgdiff} for equal disks is the double of \eqref{valorecgasint}:
\[ \frac1\kappa\left\{\sum_{eq.disks} C_{ij} - \frac{2a}{\pi}\right\} =  -\frac1{\pi^2}\log\kappa + {\cal O}(\kappa) \] 
which {\em has a logarithmic divergence}. For different disks the divergence is smoothed and the adimensional length $\kappa$ is traded for
$\beta-1$. The coefficient of the logarithmic term in the two expressions is the same, when $\beta\to 1$ the log-term in \eqref{valoreB1} transforms in the $\log\kappa$ term for equal disks. This has some importance from the practical point of view:  the unexpected universal repulsion for equal disks shows its consequences also for almost equal disks. 
From figure~\ref{figura2LogV3} it is apparent that $C_g$ follows the law of equal disks in the region $\beta-1 \ll \kappa\ll 1$:  at these distances the disks are seen as physically equal and to all effects the force grows logarithmically as $\kappa$ decreases.
 For smaller $\kappa$, i.e. $\kappa\ll \beta-1$,  the scale changes: the disks are physically different  on this scale and the logarithmic 
 divergence is traded for a constant force (in figure~\ref{figura2LogV3} there is a $\kappa$ in the denominator, i.e. in the plateau
 the variation of $C_g$ is like $B\, \kappa$ ).

\begin{figure}[!ht]
\begin{center}
\includegraphics[width=0.49\textwidth]{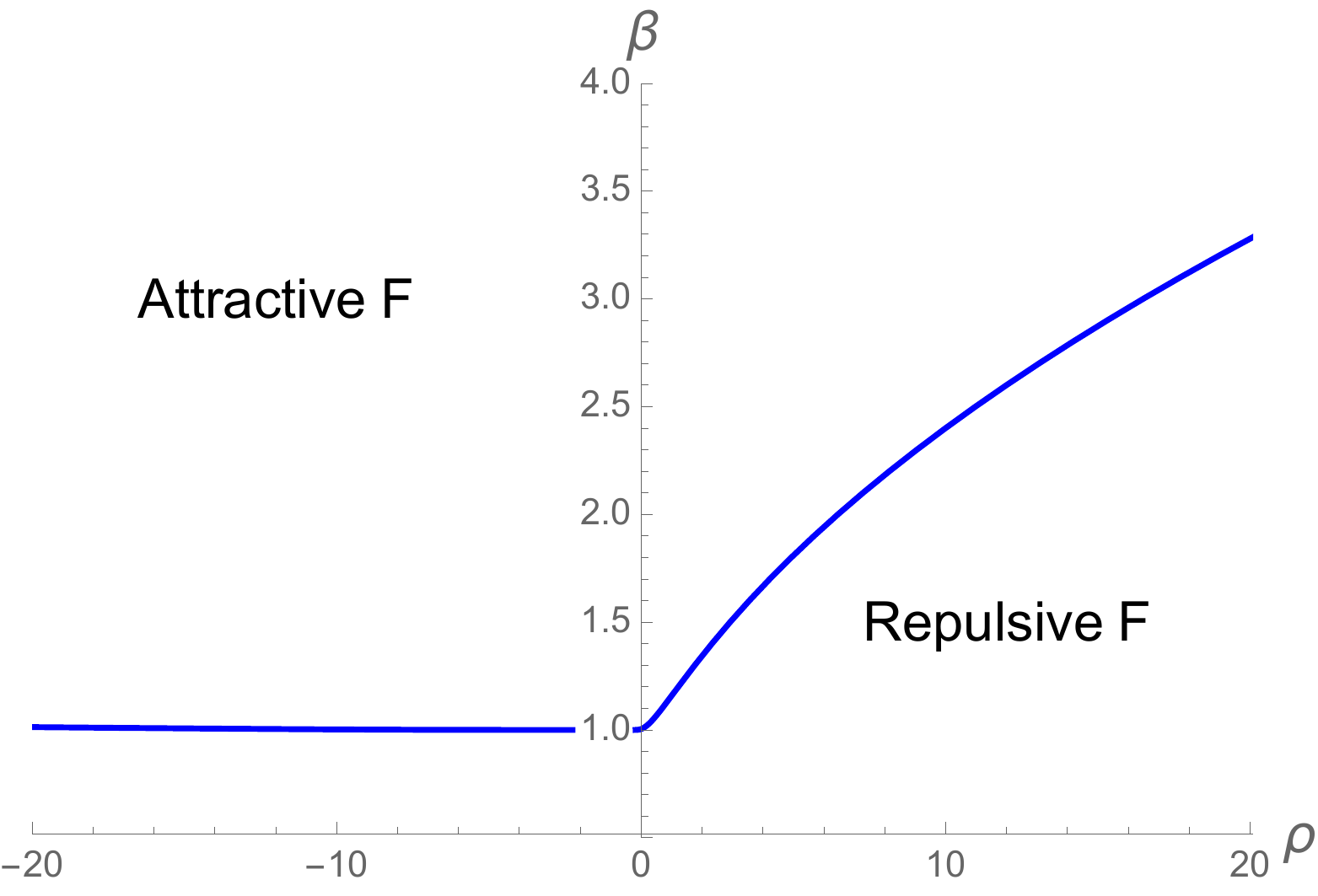}\hskip0.01\textwidth
\includegraphics[width=0.49\textwidth]{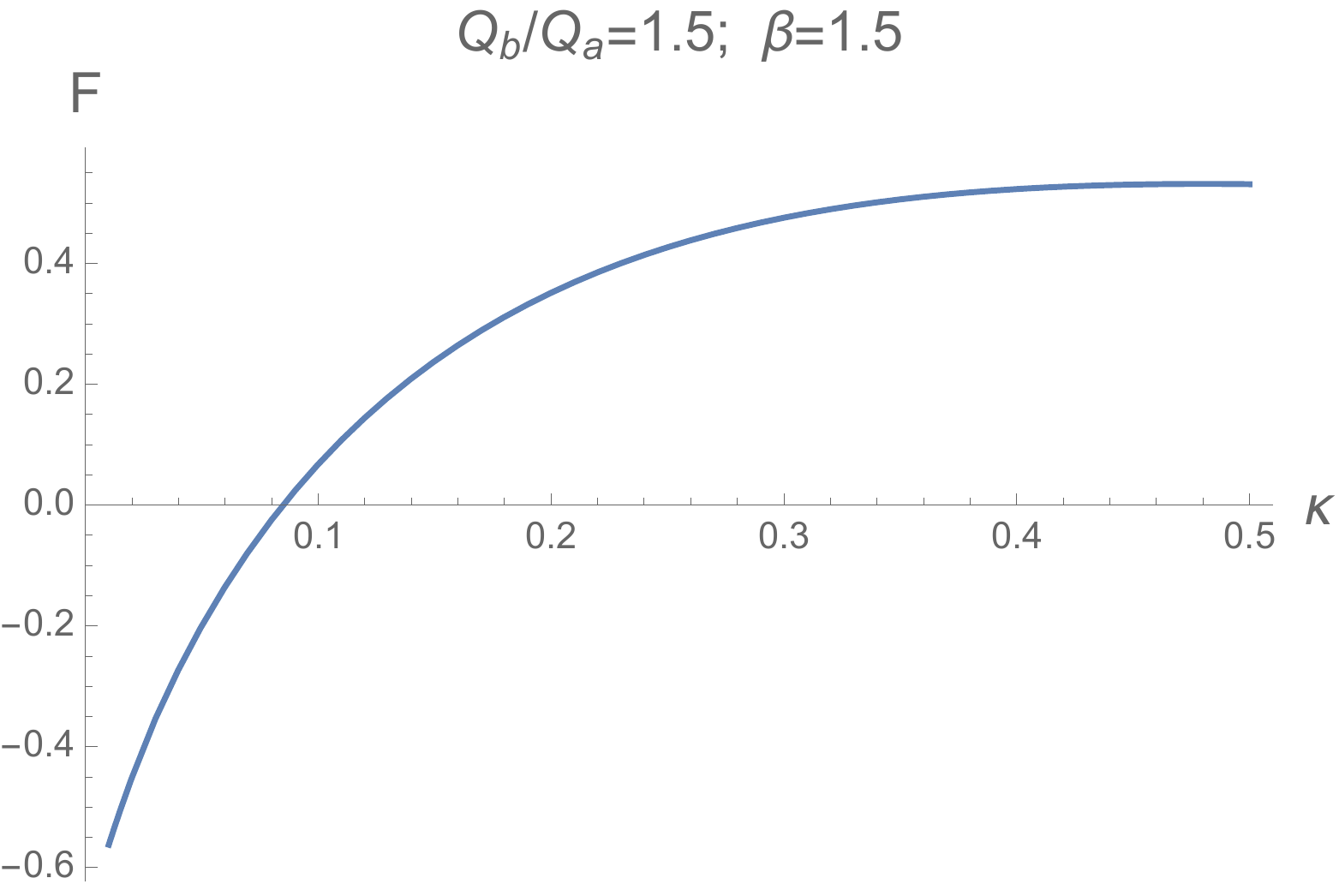}
\caption{Left panel: Attractive and repulsive domains in terms of $\beta$ and $\rho = Q_b/Q_a$. Right panel:
an example of force, in units $a=1, Q_a=1$.
\label{figbrho}}
\end{center}
\end{figure}

The analytical results, confirmed by the numerical calculations,  can be used in different circumstances. Here we give a couple of examples.
Let us consider two isolated conducting disks with charges $Q_a, Q_b$. Expanding \eqref{forzagen} for $\kappa\to0$ and using
the small distance behavior of $C_{g}$ and $C$ one easily finds\cite{mac2}
\be
 F = \frac1{a^2}\Bigl\{\frac{\pi^2}{8 \beta^2} (Q_a+Q_b)^2 B - \frac1{2\beta^2} \left( \beta(Q_a-Q_b) + \sqrt{\beta^2-1}(Q_a+Q_b)\right)^2
 \Bigr\}\,.\label{nuova53}
\ee
Using the known analytical form of $B$ as a function of $\beta$ we are now able to describe the attractive or repulsive nature of the force (at short distances) in a plane $\beta$-$\rho$, where $\rho = Q_b/Q_a$. The domains are separated by the curves $F=0$ and are shown figure~\ref{figbrho}. Let us note that for fixed $\rho\neq-1$, \eqref{valoreB1} implies that always exists a $\beta_c$ such that
for $1<\beta < \beta_c $ the first repulsive term in \eqref{nuova53} exceeds the second, attractive, term. This is in agreement with
the result \eqref{forzaQris}: the force between equal disks is always repulsive at short distances, for any charge ratio, except in the case $Q_a=-Q_b$.

An interesting feature is that for fixed $\rho$ the character the force change from repulsive to attractive 
for increasing $\beta$ 
and this can have some interest in the study of growing structures. An example of force as a function of $\kappa$ is shown in the right panel of figure~\ref{figbrho} for like charges, $\rho =1.5$. At large distances the force is repulsive, as expected, but at small
distances becomes attractive.

\begin{figure}[!ht]
\begin{center}
\includegraphics[width=0.65\textwidth]{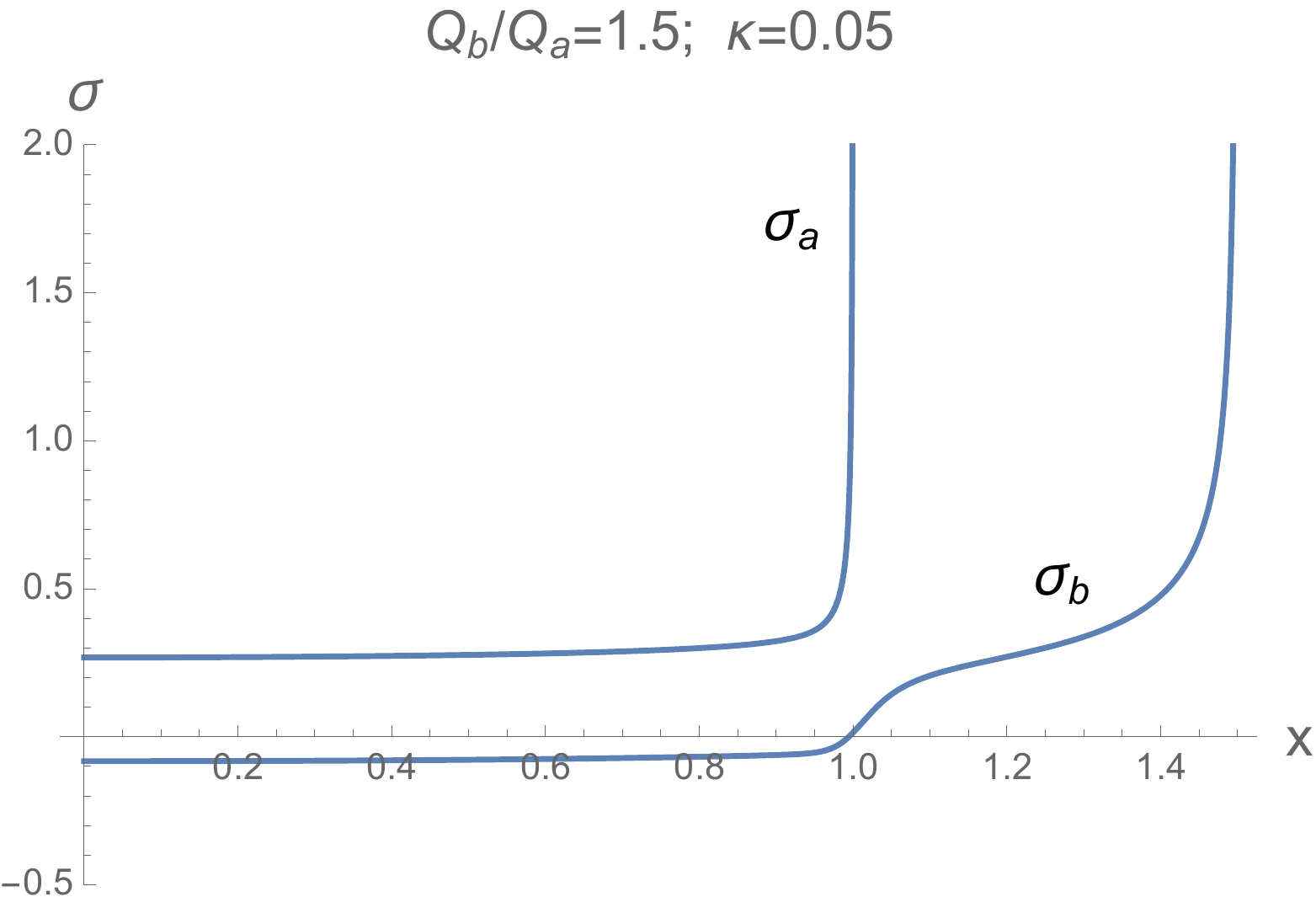}\hskip0.01\textwidth
\caption{Charge densities for two disks, with $a=1, \beta=1.5$ and like charges, $Q_b/Q_a = 1.5$, at distance $\kappa=1.5$.
\label{figsigma}}
\end{center}
\end{figure}

The reader may wonder how it is possible to have an attractive force between two disks with like charges. Polarization effects are excluded as the disks are infinitely thin. The attraction is produced by a displacement of the charges on the bigger disk.
To show this effect let us consider the case described in figure~\ref{figbrho} for $\kappa = 0.05$ and $Q_b = 1.5\, Q_a > 0$.
Solving equations \eqref{potcof2div} and computing the Abel transformation of the solutions one obtains the densities shown in 
figure~\ref{figsigma}: the larger disk acquires a {\em negative} charge density in the bulk region expelling towards the edge
the excess of positive charge. The two bulk regions, oppositely charged, but very close, attract. The outside is rejected by the small disk,
but the force is weaker, as the distances between like charged regions on the two disks are larger.

As a second example we consider the smaller disk with charge $Q_a$ and the larger disk earthed. 
This configuration is common in the elementary treatment of Kelvin microscopy.
We first note that for $\beta\to \infty$ the problem is equivalent to finding force between to equal disks with charges $\pm Q_a$ at distance $x=2\ell$, as follows by the method of images, then in this limit
\be F_\infty = -\frac{Q_a^2}2\frac{\partial}{\partial x}\frac{1}{C_{EQ}(x)} = -\frac{Q_a^2}4\frac{\partial}{\partial \ell}\frac{1}{C_{EQ}(2\ell)}\label{forza01}\ee
In the general case for $V_2=0$ we have $Q_b = Q_a C_{12}/C_{11}$ and the force can be computed, as is well known,
by taking the derivatives in \eqref{forzagen} at {\em fixed} charges and then substituting this relation. A simple calculation gives
\be F = -\frac{Q_a^2}{2a}\frac{\partial}{\partial\kappa}\frac{1}{C_{11}} \label{forza02}\ee
The same result can be obtained more simply by taking the derivative of the Legendre transformation of the usual expression of energy, as done in similar cases in textbooks.
The reader can easily verify that in effect as $C_{11}\sim 1/\kappa$ the
expansion \eqref{valoreC11def} allows the computation of $F$ up to order $o(\kappa^2)$.
Using the asymptotic expansion \eqref{valoreceq2k} we have at this order
\be F_\infty = \frac{Q_a^2}{a^2}
\left\{ - 2 + \frac{8 \kappa}{\pi}\left(\log\frac{8\pi}{\kappa}-\frac32\right)
-2\frac{\kappa^2}{\pi^2}\left( 9 \log^2\frac{8\pi}{\kappa} - 30 \log\frac{8\pi}{\kappa} +26\right)
\right\}\,.\label{forzainforder2}
\ee
In the general case \eqref{forza02} a tedious but elementary calculation gives from
\eqref{valoreC11def}:
\begin{multline}
 F = F_{\infty} - \frac{Q_a^2}{a^2}\frac{8\kappa}{\pi}\left(\arctanh\frac1\beta + 2\sqrt{\beta^2-1} - 2 \beta\right)
+ \\
\frac{Q_a^2}{a^2}\kappa^2\left[
\left(\frac{24 \left(4 \beta^2-3\right)}{\pi ^2 \sqrt{\beta^2-1}}+\frac{84 \beta-96 \beta^3}{\pi
   ^2 \left(\beta^2-1\right)}+\frac{36 \arctanh\frac1\beta}{\pi ^2}\right) 
   \log\frac{8 \pi }{\kappa} + R(\beta)
\right]
\label{forzabfinito}
\end{multline}
\begin{multline*}
R(\beta) = 24\left(Y_2(\beta)+Y_1(\beta)\right) -\frac{24}{\pi^2} \left(\arctanh\frac1\beta\right)^2 
-\frac{12}{\pi^2}(5 - 8\beta + 8\sqrt{\beta^2-1})\arctanh\frac1\beta\\
-\frac4{\pi^2 (\beta^2-1)}(48 \beta^4-32 \beta^3-72 \beta^2+31 \beta+24)
+ \frac{8}{\pi^2 \sqrt{b^2-1}}(24 \beta^3-16 \beta^2-24 \beta+15)
\end{multline*}
The finite size effects for $\beta<\infty$ are clearly displayed in \eqref{forzabfinito}.

\section{Computation of the short distance expansion\label{sezioneris}}
To study the structure of divergences in the two disks system it is very convenient to use the Hilbert space notation introduced in 
section~\ref{sezdischiuguali}. In terms of the integral operators ${\cal A}, {\cal B}$ associated to the kernels \eqref{twokernels} the solutions of equations \eqref{appb1.3} and \eqref{appb1.3g1} take the form
\be
\ket{f_1} = \dfrac{1}{1-{\cal A}+{\cal B}}\ket 1\,;\quad \ket{g_1} = - \dfrac{1}{1-{\cal A}+{\cal B}}\ket {G_\beta}
\label{eq1}
\ee
while the capacitance coefficients are:
\be
\begin{split}
&C_{11} = \frac{2a}{\pi}\bra 1 \dfrac{1}{1-{\cal A}+{\cal B}}\ket 1\,;\qquad
C_{21} = - \frac{2a}{\pi}\bra {G_\beta} \dfrac{1}{1-{\cal A}+{\cal B}}\ket 1\,;\\
&C_{12} = - \frac{2a}{\pi}\bra 1  \dfrac{1}{1-{\cal A}+{\cal B}}\ket {G_\beta}\,;\qquad
C_{22} = \frac{2 a \beta}{\pi} + \frac{2a}{\pi}\bra{G_\beta}\dfrac{1}{1-{\cal A}+{\cal B}}\ket{G_\beta}
\end{split}\label{eq2}\ee 
$\ket{G_\beta}$ represents the function defined in \eqref{valGb}. Let us note that in this language the general property $C_{12} = C_{21}$ is evident as the scalar product is hermitian.

The possibility of writing the simple set of equations \eqref{eq1} rests of the decoupling procedure: in the original form the functions $f_1, f_2$ were defined on {\it different intervals} and the construction would have been more complicated.

The operator ${\cal B}$ vanishes for $\beta\to \infty$ 
and is smoother than ${\cal A}$ for $\kappa\to 0$, as internal integrations run on the interval $s>\beta>1$.
It is natural to perform a perturbation theory 
in ${\cal B}$ and this is essentially what has been done in \cite{paf}. The solutions $f_1, g_1$ have been computed at leading order for $\kappa\to 0$ in the form
\be
f_1(t) = f_L(t,2\kappa) + \delta f_1(t)\,;\qquad g_1(t) = - f_L(t,2\kappa) + \delta g_1(t)
\label{prsaf1g1}\ee
with:
\begin{multline}
\delta f_1(t) = 
\frac1{2\pi}\dfrac{1}{\sqrt{1-t^2}}\left(t\,{\rm arctanh}\frac{t}{\beta} - {\rm arctanh}\frac1\beta\right)\\
+\frac1{4\pi}
\left[\arctan\dfrac{1+ \beta t}{\sqrt{(\beta^2-1)(1-t^2)}} + \arctan\dfrac{1- \beta t}{\sqrt{(\beta^2-1)(1-t^2)}} \right]
\label{appB2.2}
\end{multline}

\be \delta g_1(t) = - \delta f_1(t) + \frac{1}{\pi} \arctan\sqrt{\dfrac{1-t^2}{\beta^2-1}}\,.
\label{dg1val.3}\ee
$f_L(t,2\kappa)$ is the solution of Love's equation with scale $2\kappa$, as implied by the form of operator ${\cal A}$:
\be f_L(t,2k) \simeq
 \frac1{2\kappa} \sqrt{1-t^2} + (1-t^2)^{-1/2}\frac1{2\pi}\left(
1 + \log\frac{8\pi}{\kappa} - t \log\frac{1+t}{1-t}\right)+ o(1)\label{fl2}
\ee
 The corresponding ket
will be denoted by $\ket{f_L^{(2\kappa)}}$.

Summing term by term the two sets of equations \eqref{eq2} we have, using \eqref{eq1}
\be 
\begin{split}
&C_{g1} = C_{11}+ C_{21} = \frac{2a}{\pi}\bra{1- G_\beta} \dfrac{1}{1-{\cal A}+{\cal B}}\ket1 = \frac{2a}{\pi}\scalar{\delta G_\beta}{f_1}\\
&C_{g2}= C_{12}+C_{22} = \frac{2 a \beta}{\pi}
-\frac{2a}{\pi}\bra{1- G_\beta} \dfrac{1}{1-{\cal A}+{\cal B}}\ket{G_\beta} = 
\frac{2 a \beta}{\pi} + \frac{2a}{\pi}\scalar{\delta G_\beta}{g_1}
\end{split}
\label{eq4}
\ee
where $\delta G_\beta \equiv 1 - G_\beta$. The essential point is that for $\kappa\to 0$
\be \delta G_\beta(t) \equiv 1-G_\beta(t) \simeq \frac{2 \beta \kappa}{\pi}\frac1{\beta^2- t^2} + {\cal O}(\kappa^3)\label{deltagb}\ee
The integration of this function does not produce additional divergences as the integration range is $0<t<1$ and $\beta>1$.
The function provides an additional factor $\kappa$, then to compute the sub-leading terms in \eqref{eq4} it is sufficient to 
consider $f_1, g_1$ at one order less, i.e. \eqref{prsaf1g1}.

The sum of equations \eqref{eq4}, using \eqref{prsaf1g1} and \eqref{appB2.2} gives
\be C_g = C_{g1}+ C_{g2} = \frac{2 a \beta}{\pi} + \frac{2a}{\pi}\scalar{\delta G_\beta}{f_1+ g_1} =
\frac{2 a \beta}{\pi} + a B \kappa
\label{sommacgdiff}
\ee
with
\be B = \frac{4\beta}{\pi^3} \int_0^1 \frac1{\beta^2-t^2} \arctan\sqrt{\dfrac{1-t^2}{\beta^2-1}} dt\,. \label{valoreBcoeff}\ee
From \eqref{valoreBcoeff} it is easy to show that for $\beta\to 1$:
\be B \to -\frac{1}{\pi^2}\log(\beta-1) + \text{const.} \label{valoreB1}\ee
The importance of this limit for the behavior of forces has been discussed in section \ref{diffdisksnum}.

Let us consider now the quantities $C_{g1}, C_{g2}$ separately. 
A direct computation of the integrals gives, at order $\kappa$:
\be
C_{g1}=
C_{11} + C_{21} = 
\frac{2a}{\pi}\scalar{\delta G_\beta}{f_1} =
a\Bigl\{
\frac1{\pi} \left( \beta - \sqrt{\beta^2-1}\right) + \frac{\kappa}{\pi^2\sqrt{\beta^2-1}}\log \frac{8\pi}{\kappa} + \kappa X_{g1}\Bigr\}
\label{valoreasyntcg1}\ee
The constant $X_{g1}$ is given by:
\be
X_{g1} = \int_0^1 dt \frac{4}{\pi^2}\frac{\beta}{\beta^2-t^2}
\Bigl\{\delta f_1(t) +  (1-t^2)^{-1/2}\frac1{2\pi}\left(
1 - t \log\frac{1+t}{1-t}\right) \Bigr\}
\label{valorexg1}
\ee
with $\delta f_1$ defined in \eqref{appB2.2}. $X_{g1}$ is easily computed numerically.
Let us comment on the different terms in \eqref{valoreasyntcg1}. The function $\delta G_\beta$ provides a depression factor $\kappa$ then it is apparent that the leading behavior comes from the leading term in $1/\kappa$ in $f_1$. The product gives the first term in 
\eqref{valoreasyntcg1}, reproducing the known leading result \eqref{valoriCij}. For the same reasons the only source of  the $\kappa\log\kappa$ correction comes from the $\log\kappa$ term in $f_L$. The constant $X_{g_1}$ comes from the remaining finite terms in $f_1$.
We note that there is not a $\kappa\log^2\kappa$ term.

The analogous expression for $C_{g2}$ follows from \eqref{sommacgdiff}:
\be C_{g_2} = C_{12} + C_{22} = a\Bigl\{
\frac1{\pi} \left( \beta + \sqrt{\beta^2-1}\right) - \frac{\kappa}{\pi^2\sqrt{\beta^2-1}}\log\frac{8\pi}{\kappa} + \kappa(B - X_{g1})\Bigr\}
\label{valoreasyntcg2}\ee
We have found the corrections to two linear combinations of capacitance coefficients, the computation of the corrections for an arbitrary linearly independent quantity will complete the task. The simplest choice is $C_{11}$. In the notations of \eqref{eq2}:
\be C_{11} = \frac{2a}{\pi}\bra 1 \frac1{1-{\cal A}+{\cal B}}\ket1 \label{defc11conket}\ee
The idea is to do a perturbation calculation expanding in ${\cal B}$: it vanishes for $\beta\to \infty$ and for $\kappa\to 0$ all $\beta$ grater than 1 must be ``seen'' as very large for the computation of edge effects. 

Expanding in ${\cal B}$ we have
\[ C_{11} \simeq \frac{2a}{\pi}\bra1\frac1{1-{\cal A}}\ket 1 - \frac{2a}{\pi} \bra1\frac1{1-{\cal A}} {\cal B} \frac1{1-{\cal A}}\ket 1
+ \frac{2a}{\pi} \bra1\frac1{1-{\cal A}} {\cal B} \frac1{1-{\cal A}} {\cal B} \frac1{1-{\cal A}} \ket 1 +\ldots
\]
Now ${\cal A}$ is the kernel of the Love equation for {\em equal} disks at distance $2\ell$, then $(1-{\cal A})^{-1}\ket1$ is just the Love's solution for parameter $2\kappa$, $f^{(2\kappa)}_L$ and we can write
\be C_{11} = \frac{2 a}{\pi} \scalar{1}{f^{(2\kappa)}_L} - \frac{2a}{\pi}\bra{f^{(2\kappa)}_L} {\cal B}\ket{f^{(2\kappa)}_L}
+ \frac{2a}{\pi}\bra{f^{(2\kappa)}_L} {\cal B} \frac{1}{1-{\cal A}} {\cal B} \ket{f^{(2\kappa)}_L} + \ldots
\label{exprc11.1}
\ee
The first term is just the double of the capacitance for two {\em equal} disks with parameter $2\kappa$, let us denote this capacity by $C_{EQ}$ to avoid confusions. From \eqref{corrShaw}:
\be \frac{2 a}{\pi} \scalar{1}{f^{(2\kappa)}_L} = 2 C_{EQ}(2\kappa) = 
a\left\{ \frac1{4\kappa} + \frac1{2\pi}\left[\log\frac{8\pi}{\kappa} - 1\right]
+\frac1{4\pi^2} \kappa\left[ \left(\log\frac{\kappa}{8\pi}\right)^2-2\right]
\right\}
\label{valoreceq2kbis}
\ee
Both from \eqref{defc11conket} and from \eqref{exprc11.1} it follows that $C_{11}\to 2 C_{EQ}(2\kappa)$ for $\beta\to\infty$, a result obvious from the method of images. $C_{11}$ is the charge on the smaller disk held at potential 1 when the larger disk is held at potential 0. When $\beta\to \infty$ 
the problem reduces to a disk at distance $\ell$ from a plane, the image is another disk with opposite charge at distance $2\ell$
from the former.
The potential difference between the to disks is evidently $\Delta V = 2$ 
 and the charge on disk 1 is given by
\[ Q_1 = C_{EQ}(2\kappa) \Delta V = 2  C_{EQ}(2\kappa) \,.\]
From the definition of the kernel $B$ in \eqref{twokernels} the term appearing in the first order correction is:
\begin{multline}
\frac{2a}{\pi}\bra{f^{(2\kappa)}_L} {\cal B}\ket{f^{(2\kappa)}_L} =
\frac{2a}{\pi}\int_0^1 dx \int_0^1 dy\; f^{(2\kappa)}_L(x) \int_\beta^\infty ds K(x,s;\kappa) K(s,y;\kappa) f^{(2\kappa)}_L(y)\\
= \frac{2a}{\pi}\int_\beta^\infty ds \left[\int_0^1 dy K(s,y;\kappa) f^{(2\kappa)}_L(y)\right]^2
\label{integraleB1}
\end{multline}
The variable $s$ in the integral is greater than $\beta$, then to lowest order in $\kappa$ we can neglect the $\kappa^2$ factor in the denominator of the kernel $K$ and \eqref{integraleB1} simplifies in 
\be
\frac{2a}{\pi}\bra{f^{(2\kappa)}_L} {\cal B}\ket{f^{(2\kappa)}_L} \simeq
\frac{2a}{\pi}\int_\beta^\infty dx \left[\frac{\kappa}{\pi}\int_0^1 dy \left(\frac1{(x-y)^2} + \frac1{(x+y)^2}
\right) f^{(2\kappa)}_L(y)\right]^2
\label{primoordinec1}
\ee
The kernel $B$ has provided an explicit factor $\kappa^2$ which allows us to use the ``bulk'' form of $f_L$, neglecting further finite corrections in the edge zone, i.e.
\be f^{(2\kappa)}_L(t) = 
\left\{ \frac1{2\kappa} \sqrt{1-t^2} + (1-t^2)^{-1/2}\frac1{2\pi}\left(
1 + \log\frac{8\pi}{\kappa} - t \log\frac{1+t}{1-t}\right)\right\} + {o}(1)\,.\label{hutson}
\ee
The integral in $y$ in \eqref{primoordinec1} can be performed with the result
\be
\frac{1}{2} \left(\frac{x}{\sqrt{x^2-1}}-1\right) +\frac{\kappa}{2\pi} \frac{x}{\left(x^2-1\right)^{3/2}}
\left(\log \frac{8 \pi }{\kappa}+1\right) 
    +  \kappa\, F(x)\,;\label{integrale1inB}
\ee
where
\begin{align}
      & F(x) = -\frac{2}{\pi^2
   \left(x^2-1\right)^{3/2}}
    \left\{
    \frac14\sqrt{\frac{x+1}{x-1}}
    \Phi\left(-\frac{x+1}{x-1},2,\frac{1}{2}\right)
    - \frac14 \sqrt{\frac{x-1}{x+1}}\Phi \left(-\frac{x-1}{x+1},2,\frac{1}{2}\right) 
   +\frac\pi2 x\right\}\nonumber\\
 & \Phi(z,2,\frac12) = \sum_{n=0}^\infty z^n/(n+1/2)^2 \nonumber
\end{align}
$\Phi$ is known as a Lerche transcendental function.
Squaring \eqref{integrale1inB}, expanding in $\kappa$ and performing the last integral in $x$ gives for the first order contribution to $C_{11}$:
\begin{multline}
-\frac{2a}{\pi}\bra{f^{(2\kappa)}_L} {\cal B}\ket{f^{(2\kappa)}_L} = \left\{
\frac{1}{\pi} \left(\beta - \sqrt{\beta^2-1}- \frac12\arctanh\frac{1}{\beta}\right)\right.
\\
\left.
 - \frac{\kappa}{2\pi^2} \left(\frac{\beta}{\beta^2-1}-\frac{2}{\sqrt{\beta^2-1}}+\arctanh\frac{1}{\beta}\right) 
\log \frac{8 \pi }{\kappa}
   +
   \kappa\, Y_1(\beta)\right\}
   \label{correzioneC11}
 \end{multline}
with
\begin{multline}
Y_1(\beta) = \frac1{\pi^2}\left[\frac1{\sqrt{\beta^2-1}}+\frac12\arctanh\frac1\beta\right]\\
+ \frac1{4\pi^3}\dfrac{4 \beta\sqrt{\beta^2-1}-\beta^2-1}{\beta^2-1}
\left[\sqrt{\frac{\beta-1}{\beta+1}}\Phi\left(\frac{1-\beta}{1+\beta},2,\frac12\right) - \sqrt{\frac{1+\beta}{\beta-1}}
\Phi\left(\frac{1+\beta}{1-\beta},2,\frac12\right)\right]
\end{multline}
The first term in \eqref{correzioneC11} reproduce the $\beta$-dependent part 
of $C_{11}^{(0)}$
in \eqref{valoriCij}, as expected. The  rest is a $\kappa \log\kappa$ correction: the only $\kappa\log^2\kappa$ terms in $C_{11}$ are contained in $C_{EQ}$, equation \eqref{valoreceq2k}. For large $\beta$, $Y_1(\beta)\sim 1/(6\pi^2\beta^3)$.

Consider now the next orders in \eqref{exprc11.1}. If $h(t)$ is the function represented by the vector ${\cal B}\ket{f_L^{(2\kappa)}}$,
it follows that the vector $\ket\psi = 1/(1-{\cal A})\ket h$ is the solution of Love equation with righthand side $h$.
From \eqref{complove1.3} it follows that, for $\kappa\to 0$:
\be \ket\psi = \frac1{1-{\cal A}}\ket h \to \psi(t) = \frac1{2\kappa} \int_{-1}^1 {\cal L}(t,s) h(s) ds\label{eqrefpsi}\ee
${\cal A}$ is the kernel of the Love's equation with scale $2\kappa$, this has produced the additional factor 1/2 in \eqref{eqrefpsi}.
The important point is that each factor $1/(1-{\cal A})$ gives a factor $1/\kappa$ in the series while each ${\cal B}$ produces a factor $\kappa^2$. A simple power counting for $\kappa$ shows that the only term that survives at order $\kappa$ after the computed first order is the second order in ${\cal B}$, and in this term  only the leading order of $f_L^{(2\kappa)}$ contributes:
\be \ket{h} = {\cal B}\ket{f^{(2\kappa)}}\to h(t) = \int_\beta^\infty ds K(t,s;\kappa)\int_0^1 dx  K(s,x;\kappa) \frac{1}{2\kappa}\sqrt{1-x^2}\,.\label{secorderinB}\ee
The function $h(t)$ is even, as follows from the symmetry properties of the kernel $B$ (see \eqref{valoreB}).
For the kernels it is sufficient to use the approximation \eqref{primoordinec1}. The integral gives
\begin{multline}
h(t) = \kappa\left\{
- \frac1\pi\dfrac{\beta(1-t^2) + t^2\sqrt{\beta^2-1}}{(1-t^2)(\beta^2-t^2)}+ \right. \\
\left.
 \frac1{2\pi}\frac1{\sqrt{(1-t^2)^{3}}}\left[\frac\pi2 + \arctan\dfrac{2-t^2-\beta^2}{2\sqrt{(1-t^2)(\beta^2-1)}}\right]
 \right\}
 \end{multline}
 It follows
 \be \bra{f^{(2\kappa)}} {\cal B} \frac1{1-{\cal A}} {\cal B} \ket{f^{(2\kappa)}} = \kappa Y_2(\beta)
 \ee
 where
\[ Y_2(\beta) = \frac12 \int_0^1 dt \int_{-1}^1 ds {\cal L}(t,s) h(t) h(s) =  \frac14 \int_{-1}^1 
\hskip-5pt dt \int_{-1}^1\hskip-5pt ds\,{\cal L}(t,s) h(t) h(s)\,.\]
The value of $Y_2(\beta)$ can be obtained numerically.
The final result for $C_{11}$ at order $\kappa$ is then
\begin{multline}
C_{11} = 2 C_{EQ}(2\kappa) + 
a \left\{\frac1{\pi} \left(\beta - \sqrt{\beta^2-1}- \frac12\arctanh\frac{1}{\beta}\right)
\right.\\
\left.
- \frac{\kappa}{2\pi^2} \left(\frac{\beta}{\beta^2-1}-\frac{2}{\sqrt{\beta^2-1}}+\arctanh\frac{1}{\beta}\right) 
 \log \frac{8 \pi }{\kappa}
   +
   \kappa\, Y_1(\beta) + \kappa Y_2(\beta)\right\}
   \label{valoreC11defbis}
\end{multline}
The result \eqref{valoreC11def} satisfy the general
expectation $C_{11}\to 2 C_{EQ}(2\kappa)$ for $\beta\to\infty$, as can be checked performing the limit and noticing that
$Y_1(\beta)$ and $Y_2(\beta)$ both vanish for $\beta\to \infty$.
This complete our calculation, all capacitance coefficients can be obtained by linear combinations of $C_{11}, C_{g_1}, C_{g_2}$,
see \eqref{valoreCij2pert}.
   
\section{Conclusions}
In this work we give a complete calculation of the capacitance matrix 
for two circular disks
up to order ${\cal O}(\ell)$ included, where $\ell$ is the distance between electrodes.
We show that the singular terms can be organized in a meaningful hierarchy directly connected to electrostatic forces at short distances.
The analytical work is supported by an extensive numerical calculation, to our best knowledge the first performed for different disks and the more accurate for equal disks. The classification of singular terms can be extended to different geometries and this could be of some interest in practical applications. On the theoretical side the importance of the dimensionality of the near-contact zone between electrodes is shown to play a crucial role in the behavior of the forces at short distances.


\end{document}